\documentclass[twocolumn]{aastex7}
\usepackage{comment}
\usepackage{graphicx}	
\usepackage{amsmath}	
\usepackage{amssymb}	
\usepackage{bm}
\usepackage{float}

\newcommand{\myfigwidth}{%
  \if@twocolumn
    \textwidth    
  \else
    0.5\textwidth 
  \fi
}

\newcommand{\pppt}{$\text{P}{^3}$T}

\newcommand{\rin}{r_{\mathrm{in}}}
\newcommand{\rout}{r_{\mathrm{out}}}
\newcommand{\rinij}{r_{\mathrm{in},ij}}
\newcommand{\routij}{r_{\mathrm{out},ij}}
\newcommand{\rij}{r_{ij}}
\newcommand{\rapo}{r_{\mathrm{apo}}}
\newcommand{\rperi}{r_{\mathrm{peri}}}
\newcommand{\vperi}{v_{\mathrm{peri}}}
\newcommand{\tperi}{t_{\mathrm{peri}}}

\newcommand{\mave}{\langle m \rangle}

\newcommand{\rg}{r_{\mathrm{g}}}
\newcommand{\ns}{n_\mathrm{s}}
\newcommand{\eerr}{\epsilon}
\newcommand{\eerrmax}{\epsilon_{\mathrm{max}}}
\newcommand{\eerrf}{\epsilon_{\mathrm{f}}}
\newcommand{\erra}{\epsilon_{\mathrm{a}}}
\newcommand{\dts}{\Delta t_{\mathrm{soft}}}
\newcommand{\dtsmax}{\Delta t_{\mathrm{soft,max}}}
\newcommand{\dthm}{\Delta t_{\mathrm{hard,max}}}
\newcommand{\dthi}{\Delta t_{\mathrm{hard},i}}
\newcommand{\Pin}{P_{\mathrm{in}}}

\newcommand{\trh}{t_{\mathrm{rh}}}
\newcommand{\td}{t_{\mathrm{d}}}
\newcommand{\tr}{t_{\mathrm{r}}}
\newcommand{\tcr}{t_{\mathrm{c}}}
\newcommand{\rh}{r_{\mathrm{h}}}
\newcommand{\sh}{\sigma_{\mathrm{h}}}

\newcommand{\tcrh}{t_{\mathrm{ch}}}
\newcommand{\tcpu}{t_{\mathrm{cpu}}}

\begin{document}

\title{A free-fall-based switching criterion for \pppt~hybrid $N$-body methods in collisional stellar systems}

\author[orcid=0000-0001-8713-0366, gname=Long, sname=Wang]{Long Wang}
\affiliation{School of Physics and Astronomy, Sun Yat-sen University, Daxue Road, Zhuhai, 519082, People's Republic of China}
\affiliation{CSST Science Center for the Guangdong-Hong Kong-Macau Greater Bay Area, Zhuhai, 519082, China}
\email[show]{L.W.: wanglong8@sysu.edu.cn}  

\author[gname=David, sname=Hernandez]{David M. Hernandez} 
\affiliation{Department of Physics, National Taiwan Normal University, Taipei 116, Taiwan}
\affiliation{Department of Astronomy, Yale University, Kline Biology Tower, 219 Prospect St, New Haven, CT 06511, USA.}
\affiliation{Harvard-Smithsonian Center for Astrophysics, 60 Garden St., MS 51, Cambridge, MA 02138, USA}
\email{}

\author[orcid=0009-0007-2280-1254, gname=Zepeng, sname=Zheng]{Zepeng Zheng} 
\affiliation{School of Physics and Astronomy, Sun Yat-sen University, Daxue Road, Zhuhai, 519082, People's Republic of China}
\email{}

\author[gname=Wanhao, sname=Huang]{Wanhao Huang} 
\affiliation{School of Physics and Astronomy, Sun Yat-sen University, Daxue Road, Zhuhai, 519082, People's Republic of China}
\email{}

\begin{abstract}

The \pppt~scheme is a hybrid method for simulating gravitational $N$-body systems. It combines a fast particle-tree (PT) algorithm for long-range forces with a high-accuracy particle-particle (PP, direct $N$-body) solver for short-range interactions.
Preserving both PT efficiency and PP accuracy requires a robust PT-PP switching criterion.
We introduce a simple free-fall-based switching criterion for general stellar systems, alongside the commonly used velocity-dispersion-based ($\sigma$-based) criterion. Using the \textsc{petar} code with the \pppt~scheme and slow-down algorithmic regularization for binaries and higher-order multiples, we perform extensive simulations of star clusters to evaluate how each criterion affects energy conservation and binary evolution. For systems in virial equilibrium, we find that the free-fall-based criterion is generally more accurate for low-$\sigma$ or loose clusters containing binaries, whereas the $\sigma$-based criterion is better suited for high-$\sigma$ systems. Under subvirial or fractal initial conditions, both criteria struggle to maintain high energy conservation; however, the free-fall-based criterion improves as the tree timestep is reduced, whereas the $\sigma$-based degrades due to its low-accuracy treatment of two-body encounters. 

\end{abstract}



\section{Introduction} 

Star-by-star $N$-body simulations of collisional stellar systems, such as star clusters, are challenging due to their multiple timescales. ``Collisional'' refers to systems where close encounters between stars significantly influence their dynamical evolution \citep[e.g.,][]{Binney1987}. 
Two timescales control the long-term dynamical evolution of such systems.
The crossing timescale $\tcr$ is the time required for an object to cross the system.
For star clusters, the half-mass crossing time $\tcrh$ is typically $1-10$ Myr \citep{Heggie2003,PZ2010}.
For accurate simulations, the integration time step must be much shorter than $\tcr$.
The two-body relaxation timescale $\tr$ is the time interval over which significant orbital changes for an object result from interactions with surrounding objects. 
For star clusters, the half-mass relaxation time $\trh$ is about $10^2-10^4$~Myr \citep{Heggie2003,PZ2010}; and old clusters, such as globular clusters (GCs) in the Milky Way, have undergone at least one $\trh$.
Simulating a cluster's full lifetime thus requires numerous integration steps to cover several $\trh$ and resolve individual stellar motions within one $\tcrh$.
Additionally, star formation often produces many binaries, triples, and higher-order multiples \citep{Offner2023}, where orbital periods range from days to Myrs.
Resolving the evolution of the shorest-period binaries demands extremely small integration steps \citep{Hernandez2020}, making global simulations computationally intensive.

To address extreme multi-timescale challenges, several schemes have been developed.
In direct $N$-body (particle-particle; PP) methods such as the state-of-the-art code \textsc{nbody6} \citep{Aarseth2003}, block time steps are used to bridge the gap between $\tcr$ and $\tr$. 
Objects in dense regions with frequent short-range interactions use smaller time steps for accuracy, while those in sparse regions use larger steps to save computation. 
With block time steps and without considering multiples, the computational cost is about $O(N^3)$ per $\trh$ \citep{Makino1988}.

For massive stellar systems like GCs, even block-time steps remain computationally intensive. To improve efficiency, the Ahmad-Cohen neighbor scheme \citep{Ahmad1973} and hybrid methods have been developed. 
Both approaches divide interactions into long-range and short-range forces using a neighbor radius. 
Since long-range gravitational forces are much weaker than short-range ones, they can be approximated with a minor impact on overall accuracy. 
The AC neighbor scheme uses larger time steps for long-range forces while retaining direct $N$-body algorithms. 
This method is implemented in \textsc{nbody6} \citep{Aarseth2003}.

In contrast, hybrid methods employ faster approximate algorithms with a fixed time step $\dts$, such as Barnes-Hut particle-tree (PT) \citep{Barnes1986,Oshino2011,Iwasawa2015,Ishigaki2021}, particle-mesh \citep{Hockney1981} and the fast multiple method \citep{Greengard1987}, for long-range interactions. This work focuses on the hybrid \pppt~method that combines the PP and PT approaches.
Unlike the $O(N^2)$ scaling of the full force calculations in the PP method, the PT method achieves greater efficiency with an $O(N \log N)$ scaling. 
\cite{Iwasawa2016} introduced a parallelization framework for developing particle simulation codes (\textsc{fdps}) based on the \pppt~scheme, and \cite{Wang2020b} developed the \textsc{petar} code based on \textsc{fdps}. 
\textsc{petar} also incorporates the slow-down algorithm regularization (\textsc{sdar}) package to address timescale issues in multiple systems \citep{Wang2020a}. 
These codes (\textsc{nbody6} series and \textsc{petar}) have been widely used to study the dynamical evolution of star clusters.

The \pppt~method introduces a changeover region with inner and outer radii ($\rin$ and $\rout$) to smoothly transition between PP and PT treatments as objects approach each other \citep{Oshino2011}. 
The original method used a universal changeover region for all objects.
Subsequently, \cite{Wang2020b} proposed mass-dependent changeover radii for stellar systems, and \cite{Ishigaki2021} applied a variant to planetesimal systems in which the radii depend on mass, distance to the central star, and local velocity dispersion.

Properly choosing the changeover radii $\rin$ and $\rout$, and the tree timestep $\dts$, is essential for a smooth transition and physically accurate results.
Previous studies have examined how these parameters affect energy errors in star clusters \citep{Iwasawa2015,Wang2020b} and planetesimal systems \citep{Oshino2011,Iwasawa2017,Ishigaki2021}. 

For spherically symmetric clusters, \cite{Iwasawa2015} found that energy errors depend on $\theta$ and $\dts/\rout\sigma$, where $\sigma$ is the global 3D velocity dispersion. \textsc{petar} adopts this $\sigma$-based criterion to automatically set $\rout$ and $\dts$ \citep{Wang2020b}.
However, it remains unclear whether this criterion applies to more complex, non-spherical stellar systems or star clusters with binaries, as this has not been thoroughly investigated. 
Such complex systems are crucial for understanding early star cluster evolution and binary orbital dynamics. 

We present a new, simple free-fall-based criterion that can be applied for general stellar systems and compare it to existing $\sigma$-based criteria. Section~\ref{sec:method} describes the \pppt~method and the $N$-body code \textsc{petar}. Section~\ref{sec:criterion} introduces the free fall-based criterion, while Section~\ref{sec:compare} compares both criteria across isolated binaries; spherical star clusters with equal/variable masses; clusters with primordial binaries and different virial ratios; and non-spherical clumpy clusters. Section~\ref{sec:discussion} discusses limitations and potential improvements. Finally, Section~\ref{sec:conclusion} summarizes our findings.

\section{\pppt~method and $N$-body code}
\label{sec:method}

In this work, we use the \textsc{petar} code which combines the \pppt~approach with the \textsc{sdar} method for numerical simulations \citep{Wang2020b}. Specifically, \textsc{petar} combines three integration methods:
\begin{itemize}
    \item The PT method computes long-range forces using a second-order symplectic leapfrog integrator.
    \item The PP method uses a fourth-order Hermite integrator with block time steps for short-range forces on stars and centers-of-mass of multiples.
    \item The \textsc{sdar} method \citep{Wang2020b} integrates multiple systems, including hyperbolic encounters, binaries, and hierarchical few-body systems.
\end{itemize}
As our focus is the criterion for switching between the PT and PP methods, we begin by introducing the underlying mathematics and defining the relevant parameters.

The \pppt~method is done via the Hamiltonian splitting for a system with $N$ particles \citep[e.g.,][]{Oshino2011}:
\begin{equation}
  \begin{aligned}
  H_{\mathrm{hard}} =& \sum_{i=1}^N \left (\frac{p_i^2}{2 m_i} - \sum_{i<j}^N \frac{G m_i m_j}{\rij} W(\rij) \right)\\
  H_{\mathrm{soft}} =& \sum_{i=1,i<j}^N  \frac{G m_i m_j}{\rij} \left[ W(\rij)-1\right ] .\\
  \end{aligned}
  \label{eq:hsplit}
\end{equation}
where $H_{\mathrm{hard}}$ indicates the short-range interactions and kinetic energies and $H_{\mathrm{soft}}$ indicates the long-range interaction; $p_i$ and $m_i$ are, respectively, the momenta and mass of the particle $i$; $\rij$ is the separation between the $i$ and $j$ particles; $G$ is the gravitational constant; $W(\rij)$ is a changeover function to smoothly transfer pieces between $H_{\mathrm{hard}}$ and $H_{\mathrm{soft}}$.

\textsc{petar} uses a mass-dependent eighth-order polynomial changeover function for each particle.
The polynomial order significantly influences the accuracy of chaotic solutions \citep{Hernandez2019a,Hernandez2019b}. An eighth-order choice guarantees smoothness through the third derivatives of the forces in the fourth-order Hermite method. 
The corresponding potential changeover function can be described as
\begin{equation}
  W(x)  = 
  \begin{cases}
    \beta (1-2 x) & (x\le 0) \\
    \beta (1-2 x) - 1 + f(x) & (0 < x < 1)\\
    0  & (x\ge 1) \\
  \end{cases}
\end{equation}
where
\begin{equation}
  \begin{aligned}
    f(x) & = 1 + \beta x^5 \left( 14 - 28 x + 20 x^2 -5 x^3 \right) \\
    x & = \frac{\rij-\rinij}{\routij-\rinij} \\
    \beta & = \frac{\routij-\rinij}{\routij+\rinij}. \\  
  \end{aligned} 
\end{equation}
Here $\rinij$ and $\routij$ denote the inner and outer boundaries of the changeover function for each object pair $i$ and $j$. 
To ensure that the force changeover function ranges from 1 to 0, $W(x)\neq 1$ for $x\le0$, see \cite{Wang2020b} for details.

With a reference value $r_{\mathrm{in}}$ set as a simulation parameter, each object has
\begin{equation}
    r_{\mathrm{in},i} = \left(\frac{m_i}{\langle m \rangle}\right)^{1/3} r_{\mathrm{in}}
    \label{eq:rinoutij}
\end{equation}
where $m_i$ is the mass of object $i$ and $\langle m \rangle$ is the average mass. The $1/3$ exponent reflects the scaling from tidal forces. The pairwise radius is then defined as the maximum of the two radii:
\begin{equation}
    \rinij = \max \left(r_{\mathrm{in},i}, r_{\mathrm{in},j}\right).
    \label{eq:rinoutij2}
\end{equation}
The ratio $\rin/\rout$ ($\rinij/\routij$) is fixed for all objects. In our simulations, we set $\rin/\rout = 0.1$, leaving only $\rin$ to be determined for the switching criterion.

For time steps in the \pppt~method, the PT part uses a shared time step $\dts$, while the PP part assigns each particle an individual time step based on \citep{Aarseth2003,Oshino2011}:
\begin{equation}
    \begin{aligned}
    \dthi = & \min\left( \eta \sqrt{\frac{ \sqrt{|\bm{A}^{(0)}_i|^2 + A^2_{0}}\, |\bm{A}^{(2)}_i| + |\bm{A}^{(1)}_i|^2 }{ |\bm{A}^{(1)}_i|\, |\bm{A}^{(3)}_i| + |\bm{A}^{(2)}_i|^2 } }, \right.\\
            & \left. \dthm \right),
    \end{aligned}
  \label{eq:dth}
\end{equation}
where $\bm{A}^{(0)}_i$ is the acceleration of particle $i$; $\bm{A}^{(j)}_i$ is its $j$-th time derivative; $A_{0}$ is a scalar smoothing parameter (with units of acceleration) to prevent unnecessarily small steps; and $\dthm$ is the maximum time step. 
In our simulations, we set $\dthm = \dts$ to smoothly couple the PT and PP methods, adopt $\eta=0.1$ as a standard value \citep[e.g.,][]{Aarseth2003}, and define $A_{0}= G \mave/\rout^2$ as the scalar acceleration from a mass $\mave$ at distance $\rout$.

To switch on the \textsc{sdar} method for a multiple system, we select members using a mass-dependent radial criterion $\rg$.
In this work, $\rg$ is uniform for all particles and satisfies $\rg < \rin$.
The switching rule also considers velocities and external perturbations. For a two-body \textsc{sdar} system on a hyperbolic orbit, if the radial relative velocity is high enough that the predicted next-step separation exceeds $\rg$ while the current separation is $>0.2~\rg$, the system is terminated. An \textsc{sdar} system experiencing strong external perturbations is terminated. The same perturbation criterion is applied when generating an \textsc{sdar} system.

\section{Free-fall-based switching criterion}
\label{sec:criterion} 

\subsection{Criterion for two-body system}

Because both $\rin$ and $\dts$ are free parameters, a linking criterion is needed to preserve efficiency and accuracy in the \pppt~scheme.
Based on numerical experience, \cite{Iwasawa2015} proposed the $\sigma$-based criterion:
\begin{equation}
    \dts = \alpha \frac{\rin}{\sigma},
    \label{eq:sigmabase}
\end{equation}
where $\alpha$ is a user-defined coefficient.
It 
is intended for spherically symmetric stellar systems. 
A limitation is that $\sigma$ must be measured for the specified initial condition. For systems with few objects or irregular shapes, $\sigma$ is difficult to determine.

To develop a more general switching criterion, we begin by integrating a simple binary system using the \pppt~method.
In this case, the PT part functions as a lower-accuracy (second-order leapfrog) PP method, unaffected by the tree-force approximation.
To simplify, we used only the Hermite integrator in the PP part to model the binary orbit, thus avoiding the complexity of including the \textsc{sdar} method. 
With only two objects, we denote the changeover radii as $\rin$ and $\rout$ without ``$i$'' and ``$j$''.

\begin{figure}[ht]
    \centering
    \includegraphics[width=\columnwidth]{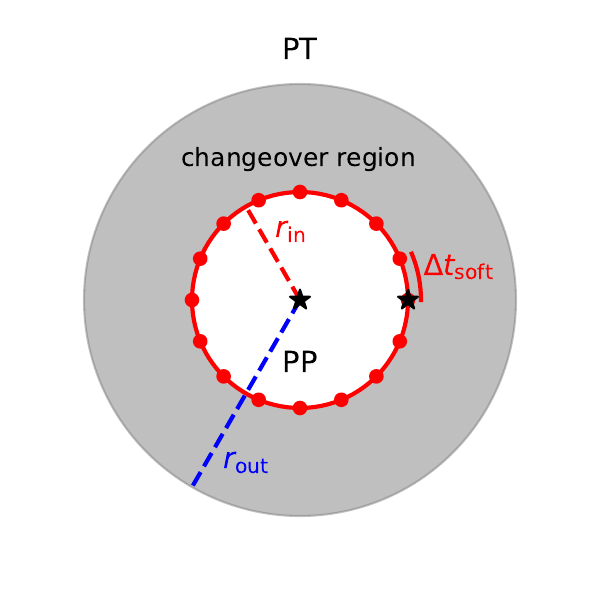}
    \caption{Free-fall-based switching criterion: set $\dts$ to $1/\ns$ of the circular binary period with semi-major axis $\rin$.}
    \label{fig:bincriterion}
\end{figure}

As demonstrated in Figure~\ref{fig:bincriterion}, if the binary orbit is within their changeover region, both the PT and PP methods are used to integrate the orbit. 
As the separation between the two components passes through the changeover region, a reasonable criterion should ensure that the PT part with a given $\dts$ matches the accuracy of the PP part.
For the PT leapfrog integrator, it is necessary to set $\dts$ much less than the binary period to ensure accuracy. 
Thus, the maximum time step $\dtsmax$ can be described as $P/\ns$, where $n_\mathrm{s}$ and $P$ are the minimum number of time steps per orbit and the binary period, respectively.
In the changeover region, a circular orbit in the inner boundary with the semi-major axis $a$ equal to $\rin$ has the shortest period $\Pin$ that requires the leapfrog method to handle. 
Thus, $\Pin/\ns$ gives $\dtsmax$ for all binary orbits crossing changeover region. The resulting switching criterion of the \pppt~method is
\begin{equation}
    \dts = \frac{\Pin}{n_\mathrm{s}} = \frac{2 \pi}{\ns}  \sqrt{\frac{\rin^{3}}{G \left(m_{1} + m_{2}\right)}},
    \label{eq:crit}
\end{equation}
where $m_1$ and $m_2$ are masses of binary components.
Using this criterion, either $\dts$ or $\rin$ can be calculated if the other is known.
Since the criterion formula resembles the free-fall time at scale $\rin$, we call it the "free-fall-based" criterion.

The above analysis considers only the circular binary case with $a=\rin$. For eccentric binaries or hyperbolic encounters with pericenter separation $\rperi=\rin$, the period or encounter timescale is longer than $\Pin$, but the velocity at $\rperi$ exceeds the circular velocity. 
Thus, achieving an integration accuracy comparable to the circular case requires a larger $\ns$.

For eccentric orbits or hyperbolic encounters, we can use the pericenter-passage timescale $\tperi$ instead of $\Pin$ to derive a new criterion.
The timescale $\tperi$ is defined as
\begin{equation}
    \tperi =  2\pi \frac{\rperi}{\vperi} = 2\pi\sqrt{\frac{\rperi^3}{G(m_1+m_2) (1+e)}}
    \label{eq:tperi}.
\end{equation}
With $\rin=\rperi$, we obtain $\tperi = \Pin/  \sqrt{1+e}$.
Therefore, replacing $\Pin$ with $\tperi$ in Equation~\ref{eq:crit} introduces only a factor of $1/\sqrt{1+e}$:
\begin{equation}
    \dts = \frac{\tperi}{n_\mathrm{s}} = \frac{2 \pi}{\ns}  \sqrt{\frac{\rin^{3}}{G \left(m_{1} + m_{2}\right) (1+e)}}.
    \label{eq:critecc}
\end{equation}
Since $\vperi \propto \sqrt{(1+e)/|1-e|}$, the largest energy error occurs near $e\sim1$. In this limit, Equation~\ref{eq:critecc} and \ref{eq:crit} differ by only a factor of $1/\sqrt{2}$, so Equation~\ref{eq:crit} is a reasonable criterion for all cases, and we adopt it for the free-fall-based criterion. However, the numerical tests in the following section show that maintaining high energy conservation at $e\sim1$ remains challenging. 

\subsection{Validation for eccentric binaries}

\begin{figure}[ht]
    \centering
    \includegraphics[width=\linewidth]{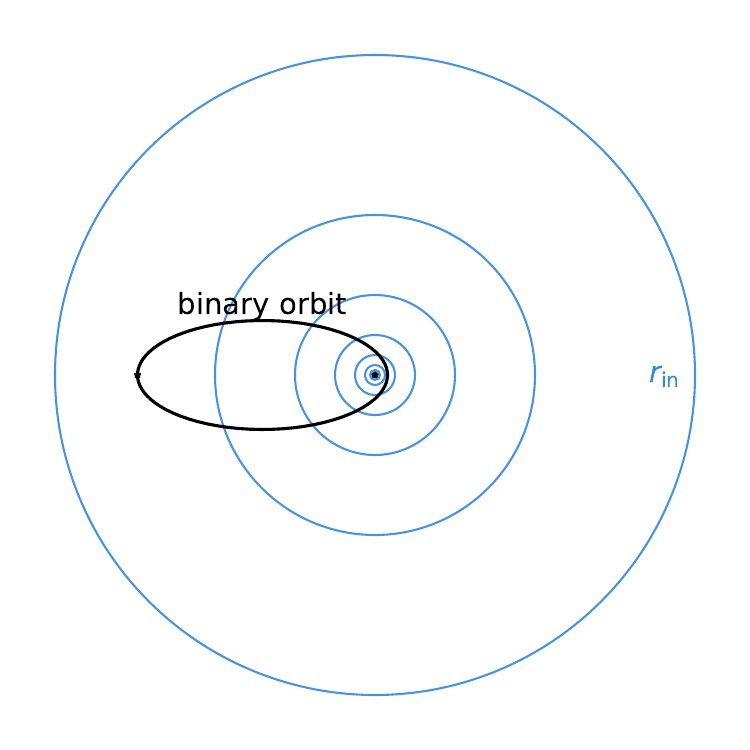}
    \caption{Eccentric binary orbits shown for various $\rin$.}
    \label{fig:binrin}
\end{figure}

To test the free-fall-based criterion and examine the effect of eccentricity, we perform two sets of simulations:
\begin{itemize}
    \item First, we fix the binary orbit and vary $\rin$ and $\dts$ to quantify integration errors, evaluate the criterion, and find an appropriate $\ns$.
    \item Second, we fix $\rperi = \rin$ and vary $\dts$ and $e$ to investigate how $\ns$ depends on $e$.
\end{itemize}

For the first set, we simulate one orbital period of binaries with semi-major axis $a=0.1$~pc, eccentricity $e=0.9$ and two mass ratios: an equal-mass case with $m_1 = m_2 = 1 \text{M}_\odot$ and an unequal-mass case with $m_1/m_2 = 0.1$ and $m_1=1 \text{M}_\odot$.

\begin{figure}[ht]
    \centering
    \includegraphics[width=\linewidth]{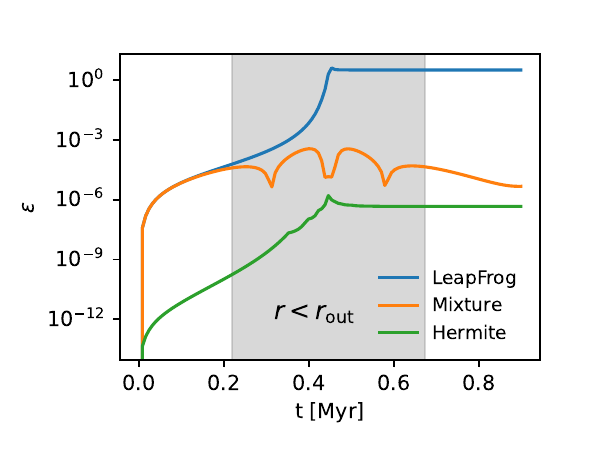}
    \caption{Relative cumulative energy error $\eerr$ for the binary with $m_1/m_2=0.1$ and $\dts=0.0078125$~Myr at three $\rin$ values: 0.0005 (pure leapfrog), 0.016 (mixture) and 0.256 (pure Hermite). Simulations start at the apocenter and reach pericenter at about 0.446 Myr. The gray band marks the interval in the mixture scheme when the integration traverses the changeover region (binary separation $r<\rout$).}
    \label{fig:eerr-threemethod}
\end{figure}

We first apply no criterion and select three values of $\rin$ with a fixed $\dts$ to examine how the relative cumulative error of energy $E$ ($\eerr = |E-E_0/E_0|$) depends on the choice of integration method for the binary with $m_1/m_2=0.1$. By adjusting $\rin$, the binary orbit can be integrated using three approaches: pure leapfrog ($\rin = 0.0005$~pc, with $\rout$ below the binary pericenter), a mixed leapfrog-Hermite (mixture) scheme ($\rin = 0.016$~pc, with the binary orbit crossing the changeover region), and pure Hermite ($\rin=0.256$~pc, with $\rin$ above the binary apocenter). 

Figure~\ref{fig:eerr-threemethod} compares these methods for $\dts=2^{-7}=0.0078125$~Myr (corresponding to about 114 steps per binary period).   
Simulations start at the apocenter and reach pericenter at about $1$~Myr. 
At peri-center, the pure leapfrog method shows a large error above $1$ and remains high.
Note that although LeafFrog is time-reversible, the error may not recover if $\dts$ is too large, as shown here.
The pure Hermite method reaches a much lower error at the pericenter of $10^{-6}$ and persists because it is not reversible. 
The mixture scheme exhibits intermediate errors: it matches the leapfrog error before entering the changeover region ($r > \rout$), oscillates inside, and partically recovers after leaving.
This behavior is consistent with expectations for the \pppt~scheme.

In general, the choices of $\dts$ and $\rin$ influence the maximum pericenter error $\eerrmax$. 
Increasing $\dts$ leads to higher $\eerrmax$ for leapfrog, while Hermite with adaptive time steps is less sensitive. 
By increasing $\rin$ (and $\rout$), the large leapfrog errors near pericenter can be avoided. 
With an appropriate combination of $\dts$ and $\rin$ ($\rout$) in the mixture approach, as shown in Figure~\ref{fig:eerr-threemethod}, the error at the end can match that of pure Hermite.
Using Equation~\ref{eq:crit}, the corresponding $\ns\approx 17$.
\cite{Hernandez2020} found that the relevant metric for accuracy was the number of steps per effective period at pericenter, which needs at least about $16$ steps for a Wisdom-Holman integrator, similar to the $\ns$ found here for the leapfrog-Hermite mixture.

\begin{figure}[ht]
    \centering
    \includegraphics[width=\linewidth]{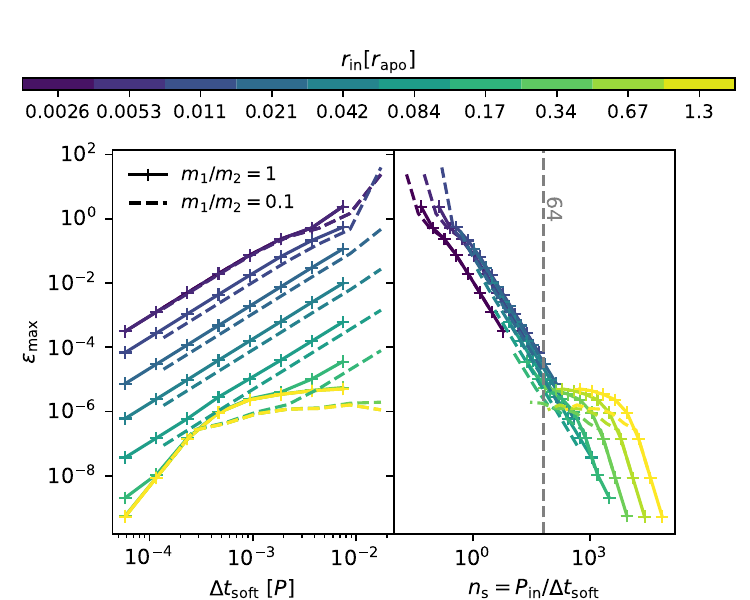}
    \caption{The maximum error of energy ($\eerrmax$) is shown as a function of $\dts$ (left) and $\ns$ (right). Colors represent different $\rin$ in units of apo-center separation $\rapo$, while line styles indicate the two binary mass ratios.}
    \label{fig:dtrinbinary}
\end{figure}

By running simulations with nested combinations of $\dts$ and $\rin$ for the same binaries, we examine how $\eerrmax$ depends on $\dts$ and $\rin$  (left panel of Figure~\ref{fig:dtrinbinary}) and on $\ns$ from the free-fall-based criterion in Equation~\ref{eq:crit} (right panel).
The inner changeover radius, $\rin$, ranges from $0.0005$~pc to $0.256$~pc, doubling at each step (see Figure~\ref{fig:binrin}), with the minimum and maximum values corresponding to pure leapfrog and Hermite integration, respectively. 
Similarly, the time step, $\dts$, doubles from $2^{-13}$~Myr to $2^{-5}$~Myr. 

The left panel of Figure~\ref{fig:dtrinbinary} shows that in the mixture approach with $\rin \ll \rapo$, $\eerrmax$ scales as $O(\dts^2)$, consistent with the second-order leapfrog method. 
Lower mass ratios result in larger errors. 
For $\rin > \rapo$, pure Hermite integration (yellow) with adaptive time steps (Equation~\ref{eq:dth}) makes $\eerrmax$ independent of $\dts$ for $\dts > 3\times 10^{-4}~P$, where $P$ is the binary period. 
For smaller $\dts$, the adaptive time step estimated from the left side of Equation~\ref{eq:dth} exceeds $\dts$ ($\dthm$), so the Hermite step is fixed at $\dts$, and $\eerrmax$ scales as $O(\dts^4)$, consistent with the fourth order.

For a fixed $\dts$, $\eerrmax$ decreases as $\rin$ decreases.
In contrast, when plotted against $\ns$, the $\eerrmax$ curves for different $\rin$ and $m_1/m_2$ almost overlap (see the right panel of Figure~\ref{fig:eerr-threemethod}).
This indicates that simulations with varying $\dts$ and $\rin$ at a fixed $\ns$ produce similar $\eerrmax$, supporting free-fall as a suitable criterion to infer $\dts$ or $\rin$ from each other with a guaranteed $\eerrmax$. 
A vertical line at $\ns = 64$ marks where $\eerrmax$ for the \pppt~scheme matches that of the pure Hermite case ($\rin = 1.3~\rapo$). When $\ns$ is smaller, $\eerrmax$ is dominated by the leapfrog error. 

For a fixed $\dts/P$, the dependence of $\eerrmax$ on $m_1/m_2$ varies with the integration method. In the pure leapfrog case with the smallest $\rin$, $\eerrmax$ is independent of $m_1/m_2$. In contrast, for the pure Hermite scheme with the largest $\rin$, $\eerrmax$ becomes gradually larger for $m_1/m_2=0.1$ when $\dts>3\times10^{-4}~P$, as the adaptive time-step method controls the actual integration steps. For smaller $\dts$, where the adaptive step is suppressed, $\eerrmax$ no longer depends on $m_1/m_2$, similar to leapfrog.  
For all other $\rin$, $\eerrmax$ depends on $m_1/m_2$ due to the Hermite contribution.  
These results indicate that the integrator yields consistent errors when the number of time steps per orbit is fixed. The higher $\eerrmax$ for the Hermite case with adaptive stepping arises because the adaptive scheme of Equation~\ref{eq:dth} does not depend on the component masses, giving identical time steps for different $m_1/m_2$. Since $P$ varies with $m_1/m_2$, $\dts/P$ is relatively larger for $m_1/m_2=0.1$, leading to a lower $\eerrmax$. 

\begin{figure}[ht]
    \centering
    \includegraphics[width=\linewidth]{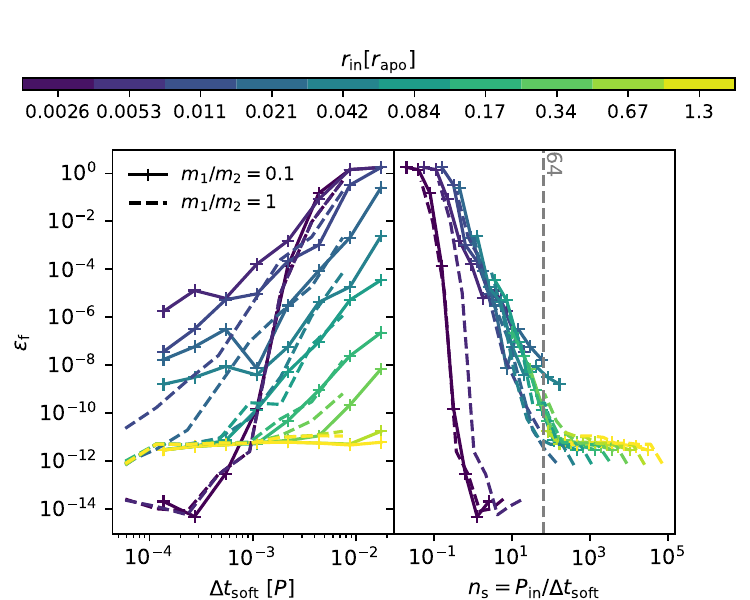}
    \caption{Similar to Figure~\ref{fig:dtrinbinary}, but replacing $\eerrmax$ by $\eerrf$.}
    \label{fig:efdtrinbinary}
\end{figure}

When plotting the final relative error $\eerrf$ instead of $\eerrmax$, the behavior differs (Figure~\ref{fig:efdtrinbinary}). $\eerrf$ is systematically lower than $\eerrmax$ and shows a weaker second-order leapfrog scaling with $\dts$. 
In the pure leapfrog case ($\rout =0.5~\rperi$; purple), the error is near $10^{-14}$ (roundoff) for $\dts< 3\times10^{-4}~P$, but jumps to $1$ for $\dts > 10^{-2}~P$. The low error at small $\dts$ stems from leapfrog's time-revisibility: errors peak at pericenter and then return back. For large $\dts$, the error does not recover, as illustrated in Figure~\ref{fig:eerr-threemethod}.

Using the previously adopted $\eta=0.1$ in Equation~\ref{eq:dth}, the Hermite error is of order $10^{-6}$, which is relatively large because the method is not time-reversible. We therefore reduce $\eta$ to 0.01 to achieve an error level of $10^{-11}-10^{-12}$, allowing a clearer demonstration of how the error transfer depends on $\rin$.
With this choice of $\eta$, the right panel of Figure~\ref{fig:efdtrinbinary} shows that the $\eerrf-\ns$ curve again indicates  $\ns\approx 64$ is needed to obtain  comparable errors between the mixture and pure Hermite methods. Note that in practice, this choice of $\eta$ is computationally expensive, so in the following analysis we continue to adopt the classical value of 0.1.

\begin{figure}
    \centering
    \includegraphics[width=\linewidth]{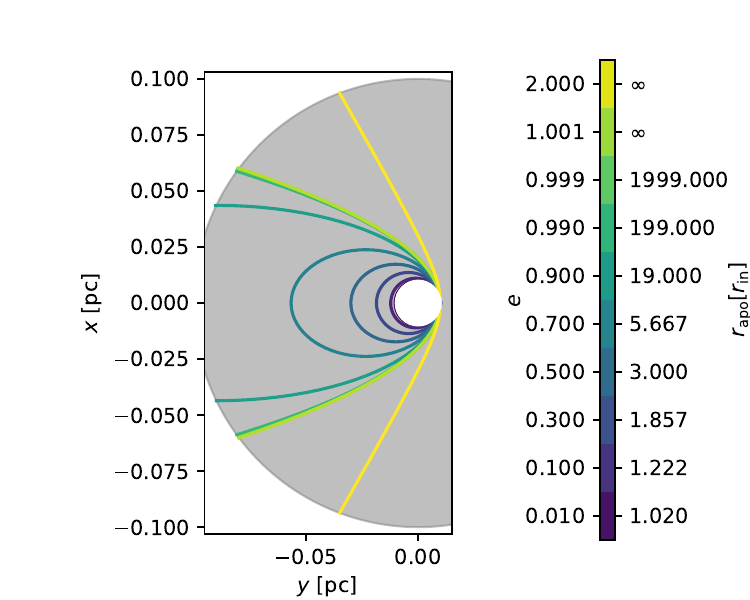}
    \caption{Orbits with various $e$ shown within the changeover (grey) region. The cases with $e > 1$ and $\rapo = \infty$ represent hyperbolic encounters.}
    \label{fig:eccorbit}
\end{figure}

The $\ns$ of 64 for $\eerrmax$ and 16 for $\eerrf$ in the above case is measured for $e = 0.9$. In the second simulation set, we examine how varying $e$ for binaries and hyperbolic encounters affects $\ns$. In all cases, $\rperi = \rin = 0.01$~pc, and the two mass ratios, 1 and 0.1, as in the first simulation set are used. Figure~\ref{fig:eccorbit} illustrates orbits for different $e$, including two hyperbolic cases with $e > 1$.

\begin{figure}
    \centering
    \includegraphics[width=\linewidth]{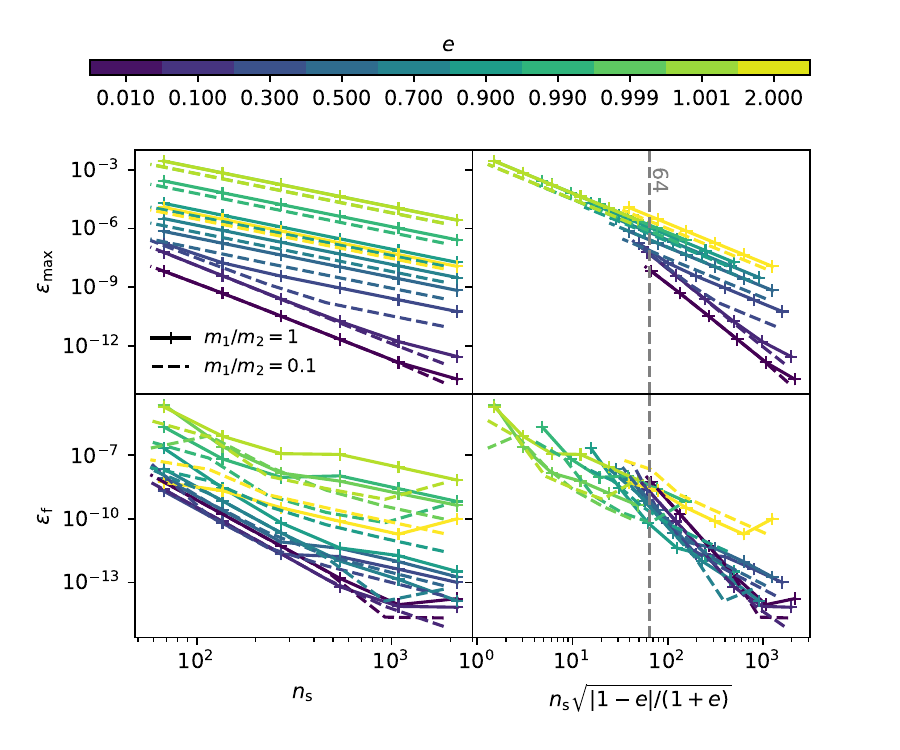}
    \caption{$\eerrmax$ and $\eerrf$ versus $\ns$ and $\ns \sqrt{|1-e|/(1+e)}$ across different eccentricties $e$ and $m_1/m_2$, with $\rperi=\rin$.}
    \label{fig:eerrmax-dts-ecc}
\end{figure}

The left panels of Figure~\ref{fig:eerrmax-dts-ecc} plot $\eerrmax$ and $\eerrf$ versus $\ns$. 
For bound binaries ($e<1$), both errors increase with $e$ at fixed $\ns$, and $\eerrf$ is less sensitive than $\eerrmax$. for hyperbolic orbits, $e=2$ yields smaller errors than $e=1.001$.
The strong $e$-dependence of the errors cannot be explained by applying the correction of $1/\sqrt{1+e}$ from Equation~\ref{eq:critecc} based on $\tperi$.
Empirically, an additional factor of $\sqrt{(|1-e|)}$ (scaling $\ns$ by $\sqrt{|1-e|/(1+e)}$) is needed to collapse the $\eerrmax$ curves, except at low $e$, as shown in the right panel of Figure~\ref{fig:eerrmax-dts-ecc}.
Thus, the free-fall-based criterion struggles to achieve low $\eerrmax$ when $e\sim1$.
Owing to the time reversibility of leapfrog, $\eerrf$ is much smaller. 
Therefore, while the free-fall-based criterion cannot guarantee uniform accuracy across all eccentricities, it remains a reasonable guideline.

\subsection{Multi-mass criterion for $N$-body systems}

Ignoring the environment perturbations, a pair force between nearby objects $i$ and $j$ in an $N$-body system like star clusters can be approximated as a Keplerian orbit. 
Thus, the free-fall-based switching criterion for \pppt~method can be extended to $N$-body systems.

In an $N$-body system, each object may have a different mass and changeover radii.
For the free-fall-based criterion in Equation~\ref{eq:crit}, we consider a pair of objects with masses $m_i$ and $m_j$ and replace $\rin$ with $\rinij$ (Equation~\ref{eq:rinoutij} and \ref{eq:rinoutij2}), yielding
\begin{equation}
    \begin{aligned}
    \dts =  \frac{\Pin}{n_\mathrm{s}} = & \frac{2 \pi}{\ns}  \sqrt{\frac{\max{(m_i,m_j)}}{\mave}\frac{\rin^{3}}{G \left(m_{i} + m_{j}\right)}} \\
           \ge &  \frac{2 \pi}{\ns}  \sqrt{\frac{\rin^{3}}{2 G \mave}}
    \end{aligned}
    \label{eq:imfcrit}
\end{equation}
In particular, $\dts$ reaches its maximum value when $m_i = m_j$, becoming independent of the individual masses. We therefore adopt this maximum value as the universal criterion for the entire stellar system.

\section{Comparing two criteria for $N$-body systems}
\label{sec:compare}

\subsection{Estimating $\rin$ from $\dts$}

\begin{figure}
    \centering
    \includegraphics[width=\linewidth]{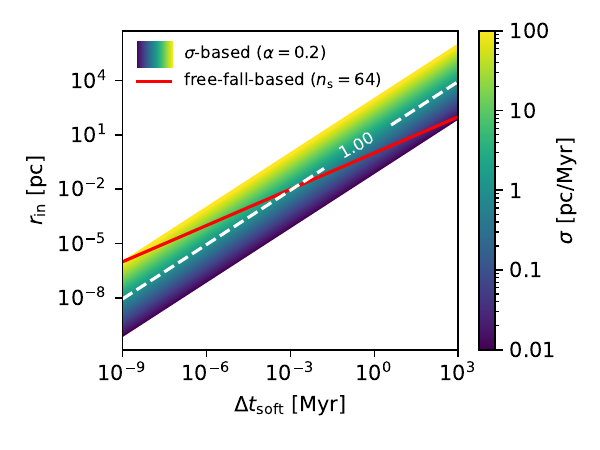}
    \caption{$\rin$ calculated from $\dts$ using the $\sigma$-based (color region) and free-fall-based criteria (red line) over a range of $\sigma$.}
    \label{fig:dts-rin-criteria-compare}
\end{figure}

We compare the estimated $\rin$ as a function of $\dts$ for the $\sigma$-based and free-fall-based criteria in $N$-body systems, using only initial conditions and no simulations, as shown in Figure~\ref{fig:dts-rin-criteria-compare}.
The systems span a wide range of velocity dispersions, $\sigma=0.01-100$~pc/Myr, representing environments from loose stellar associations to dense nuclear star clusters. 
We adopt $\ns=64$ for the free-fall-based criterion and $\alpha=0.2$ for the $\sigma$-based criterion.
These parameter choices are also used in subsequent comparisons for different stellar systems.

The free-fall-based criterion depending only on the object mass ($1~M_\odot$ in the plot), produces a single line independent of $\sigma$, while the $\sigma$-based criterion produces a range of $\rin$ values for a given $\dts$. This suggests that each criterion is best suited to specific stellar systems.

In typical open clusters ($\sigma\sim1~$pc/Myr), the two criteria intersect at $\dts \approx 10^{-3}$~Myr. Below this point, the free-fall-based criterion produces larger $\rin$ for the same $\dts$, leading to higher accuracy in the \pppt~method.

In low-$\sigma$ systems, such as field stars or stellar associations ($\sigma\sim0.01~$pc/Myr), the free-fall-based criterion always produces larger $\rin$ over $\dts$ values from $10^{-9}$ to $10^3$~Myr. This can improve the accuracy of binary integration (see Section~\ref{sec:binary}).

For high-$\sigma$ systems like dense nuclear star clusters ($\sigma\sim100$~pc/Myr), the $\sigma$-based criterion gives larger $\rin$, indicating it is more suitable.

The two criteria probe different aspects of stellar dynamics. From Equation~\ref{eq:sigmabase}, the $\sigma$-based criterion gives $\rin/\dts \propto \sigma \propto \sqrt{G M/\rh}$, whereas from Equation~\ref{eq:imfcrit}, the free-fall-based criterion yields $\rin/\dts\propto \sqrt{G\mave/\rin}$. Thus, the $\sigma$-based criterion is sensitive to the global gravitational potential, while the free-fall-based criterion is sensitive to the mutual gravity within $\rin$. 
Figure~\ref{fig:dts-rin-criteria-compare} further indicates that in low-$\sigma$ systems the potential of a star is dominated by mutual gravity, whereas in high-$\sigma$ systems, it is dominated by the global potential.

\subsection{Spherical star clusters with equal-mass objects}
\label{sec:equalmass}

\subsubsection{Energy error depending on $\dts$}
\label{sec:errdt}

We perform $N$-body simulations of the spherical cluster model with $N=1000$, $10000$ and 100000, and $\rh=1$~pc, consisting of equal-mass objects in virial equilibrium, applying both criteria to assess their impact on integration accuracy as $\dts$ or $\rin$ varies. 
To eliminate the effect of opening angles in the PT part, we set $\theta=0$.
To avoid large errors during close encounters and binary formation, we switch to the \textsc{sdar} method using the criterion parameter $\rg=10^{-4}$~pc. This setting applies to all star cluster simulations in this work, except those with primordial binaries.
For $N=1000$, five sets of initial conditions with different random seeds are used to capture statistical variation, while a single set is used for the larger $N$ values.
To ensure most objects complete at least one orbit, simulations run until two half-mass crossing times, where
\begin{equation}
  \tcrh = \sqrt{\frac{\rh^{3}}{G M}},
\end{equation}
and $M$ is the total cluster mass.

\begin{figure}
    \centering
    \includegraphics[width=\linewidth]{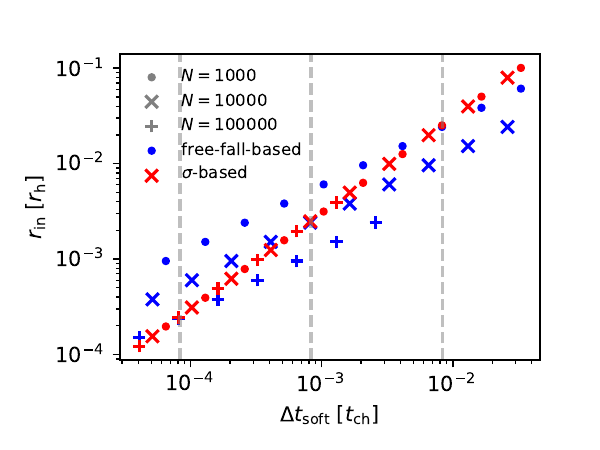}
    \caption{$\rin$ estimated from $\dts$ by two criteria for the spherical cluster model with $\rh=1$~pc and three different $N$. Vertical lines show the transition $\dts$.}
    \label{fig:rin-dts-equalmass-rh1}
\end{figure}

We vary $\dts$ from $2^{-17}$ to $2^{-6}$~Myr in logarithmic steps and calculate $\rin$ using Equations~\ref{eq:sigmabase} and~\ref{eq:crit} (Figure~\ref{fig:rin-dts-equalmass-rh1}) to compare accuracy over the same time interval. 
For each $N$, we use only a subset of $\dts$ values, since $\tcrh$ depends on $N$ and extremely large or small $\dts/\tcrh$ are computationally expensive. 
For example, for $N=1000$ we adopt $2^{-15}-2^{-6}$~Myr, 
corresponding to $\sim6.5\times10^{-5}-3.3\times 10^{-2}~\tcrh$.

Figure~\ref{fig:rin-dts-equalmass-rh1} shows that, when $\dts$ is expressed in units of $\tcrh$, the $\sigma$-based criterion yields the same $\rin$-$\dts$ relation for all $N$, whereas the free-fall-based criterion depends on $N$, in contrast to Figure~\ref{fig:dts-rin-criteria-compare}.
This arises because, in a virialized equal-mass system, $\sigma\propto \rh/\tcrh$ and $\rh$ is identical in all simulations, so the $\sigma$-based criterion implies $\dts[\tcrh] \propto \rin$, producing a linear relation.

The two criteria give comparable $\rin$ at different $\dts$, and this transition $\dts$ scales linearly with $N^{-1}$. For $N=1000$, 10000 and 100000, the transition occurs at $\dts \approx 8.3 \times 10^{-3}$, $8.3\times10^{-4}$ and $8.3\times10^{-5}~\tcrh$, respectively.

This relation can be understood by examining how \(\rin\) depends on \(\dts\) for the two criteria. From Equation~\ref{eq:crit}, \(\rin \propto \dts^{2/3} \mave^{1/3}\) for the free-fall-based criterion, where \(\mave = m_1 = m_2\), and from Equation~\ref{eq:sigmabase}, \(\rin \propto \sigma \dts\) for the $\sigma$-based criterion. When the two criteria yield the same \(\rin/\rh\) and \(\dts/\tcrh\), we obtain the transition timestep scales as \(\dts/\tcrh \propto \mave / (\sigma^3 \tcrh) \propto \mave (\rh/M)^{3/2} (M/\rh^3)^{1/2} \propto 1 / N\).

\begin{figure}
    \centering
    \includegraphics[width=\linewidth]{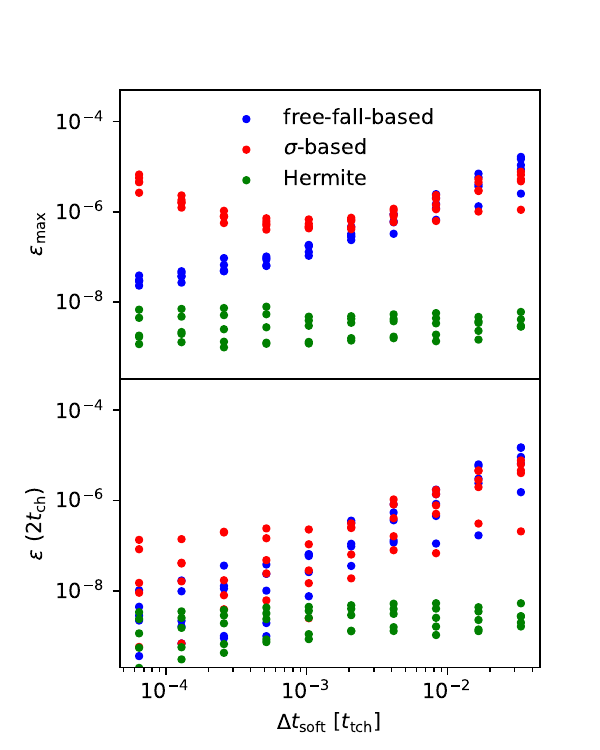}
    \caption{The maximum relative cumulative energy error $\eerrmax$ up to $2~\tcrh$ (upper panel) and the final error at $2~\tcrh$ for spherical cluster simulations with $N=1000$, $\rh=1$~pc and equal-mass objects using $\sigma$-based, free-fall-based criteria and pure Hermite method. Each parameter set is run five times with different random seeds to show statistical variation.}
    \label{fig:eerrmax-dts-equalmass-rh1}
\end{figure}

The upper panel of Figure~\ref{fig:eerrmax-dts-equalmass-rh1} shows $\eerrmax$ for all simulations with $N=1000$ using both criteria and pure Hermite method. 
Above the transition $\dts$ ($\approx 8.3\times 10^{-3}~\tcrh$), both criteria yield similarly increasing errors as $\dts$ grows. 
For larger $\rin$, the $\sigma$-based criterion gives slightly smaller errors. 
Below the transition $\dts$, the free-fall-based criterion shows an approximately linear decrease in $\eerrmax$, while the $\sigma$-based criterion exhibits increasing $\eerrmax$ as $\dts$ decreases. 

When comparing the final error at $2~\tcrh$, the trend of increasing error with decreasing $\dts$ is no longer observed for the $\sigma$-based criterion. 
In general, the free-fall-based criterion achieves better accuracy for small $\dts$, though its errors remain larger than those of the pure Hermite method, which is independent of $\dts$ due to its adaptive time-stepping but more computationally expensive.

\begin{figure}
    \centering
    \includegraphics[width=\linewidth]{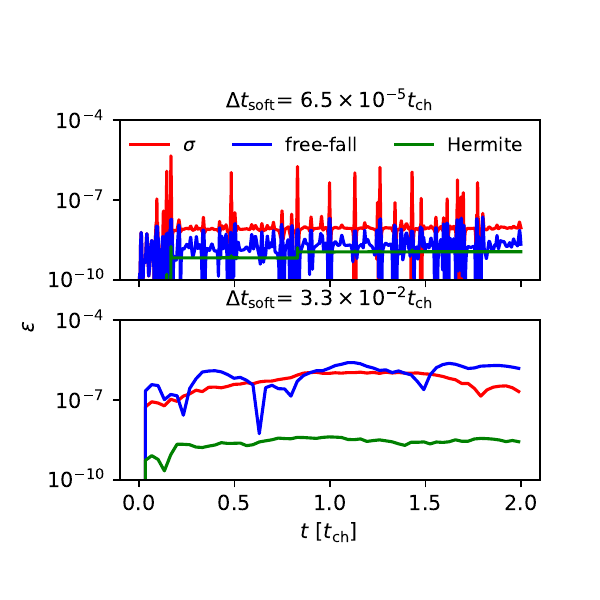}
    \caption{Relative cumulative energy error $\eerr$ over time for spherical cluster simulations with $\rh=1$~pc and equal-mass objects using the $\sigma$-based criterion, the free-fall-based criterion and the pure Hermite method. For each case, simulations with the minimum and maximum $\dts$ are compared.}
    \label{fig:eerr-t-equalmass-rh1}
\end{figure}

\begin{table}[]
    \centering
    \caption{Comparison of simulation parameters for the $\eerrmax$ test with the $\sigma$-based criterion between this work and \cite{Iwasawa2015}.}
    \label{tab:iwasawacompare}
    \begin{tabular}{c|cc}
    \hline
      $ Model$  & \cite{Iwasawa2015} & This work\\
    \hline
      $A_0$     & $0.1G\mave/\rout^2$ & $G\mave/\rout^2$ \\
      $\dthm$   & $\dts/4$           & $\dts$ \\
      N         & 131072     & 1000 \\
      softening & $4/N$              & 0 \\
      $\theta$  & 0.2, 0.4, 0.8      & 0 \\
    \hline
    \end{tabular}
\end{table}

The increase in $\eerrmax$ with decreasing $\dts$ for $\sigma$-based criterion was not seen in the benchmark by \cite{Iwasawa2015}. 
Their Figure 2 shows that varying $\rin$ yields similar $\eerrmax$ values as a function of  $\dts/\rout\sigma$. 
As summarized in Table~\ref{tab:iwasawacompare}, their simulations use several different parameters, most notably the softening parameter. 
Their $N$-body code lacks an accurate integrator for close encounters and binaries, so they apply Plummer softening to the gravitational potential to limit large $\eerrmax$. 
In contrast, we use the exact Newtonian potential without softening; thus, when $\rin$ is small, close encounters or binaries with separations outside the changeover radii can induce large errors.

Figure~\ref{fig:eerr-t-equalmass-rh1} illustrates this effect by comparing $\eerr$ evolution for the minimum $\dts$ ($\approx3\times3.3^{-2}~\tcrh$) under both criteria and the pure Hermite method. 
Both criteria show curves with occasional spikes at the same times, 
but the free-fall-based approach maintains a lower average $\eerr$ and significantly smaller spikes than the $\sigma$-based case.
These spikes correspond to brief drops in accuracy during close encounters, as errors increase when objects approach and decrease as they separate reflecting the reversible nature of the leapfrog method. 
As shown in Figure~\ref{fig:rin-dts-equalmass-rh1}, the $\sigma$-based criterion gives lower $\rin$ for small $\dts$, suggesting that the leapfrog error contributes more during close encounters.
With a larger $\rin$, as in the free-fall-based method, the Hermite scheme more accurately resolves close encounters, and suppresses large error spikes.

Figure~\ref{fig:eerr-t-equalmass-rh1} also compares $\eerr$ for the largest $\dts$ ($\approx1.6\times10^{-2}$ Myr). 
Both criteria show similar patterns, indicating poor performance at large $\dts$.
In practice, such large time steps are not used, as most objects are integrated with the Hermite method and $\eerr$ is dominated by leapfrog error, making the \pppt~method unnecessary.

\begin{figure}
    \centering
    \includegraphics[width=\linewidth]{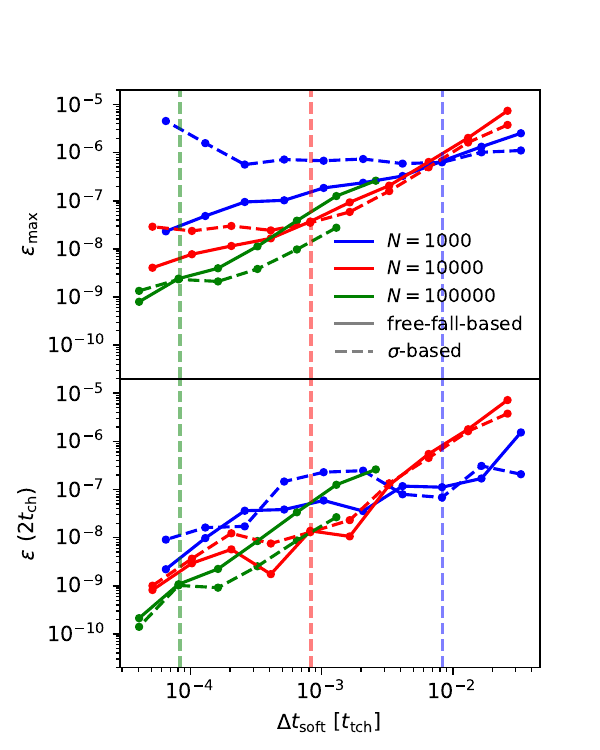}
    \caption{Errors as functions of $\dts$ for equal-mass star clusters with $\rh=1$~pc and varying $N$ under the two criteria. The upper and lower panels show $\eerrmax$ and the final error at $2~\tcrh$, respectively. Vertical dashed lines mark the transition $\dts$.}
    \label{fig:eerr-dts-equalmass-varyN}
\end{figure}

Figure~\ref{fig:eerr-dts-equalmass-varyN} compares $\eerrmax$ and $\epsilon(2\tcrh)$ for different $N$. In all cases, $\eerrmax$ shows the same general trend: below the transition $\dts$, the free-fall-based criterion performs better, whereas above the transition the $\sigma$-based criterion is superior. For $N=10000$, the difference between the two criteria is less pronounced than for $N=1000$, and the $\sigma$-based does not exhibit a strong increase in error as $\dts$ decreases below the transition, likely because mutual gravity is less dominant. For $N=100000$, the transition $\dts$ is so small that the $\sigma$-based criterion yields low errors over most of the sampled range. These results indicate that the $\sigma$-based criterion is indeed more accurate for high-$\sigma$ systems. For $\epsilon(2\tcrh)$, the error behavior is more stochastic for all $N$. 

\subsubsection{Influence of opening angles}

\begin{figure}
    \centering
    \includegraphics[width=\linewidth]{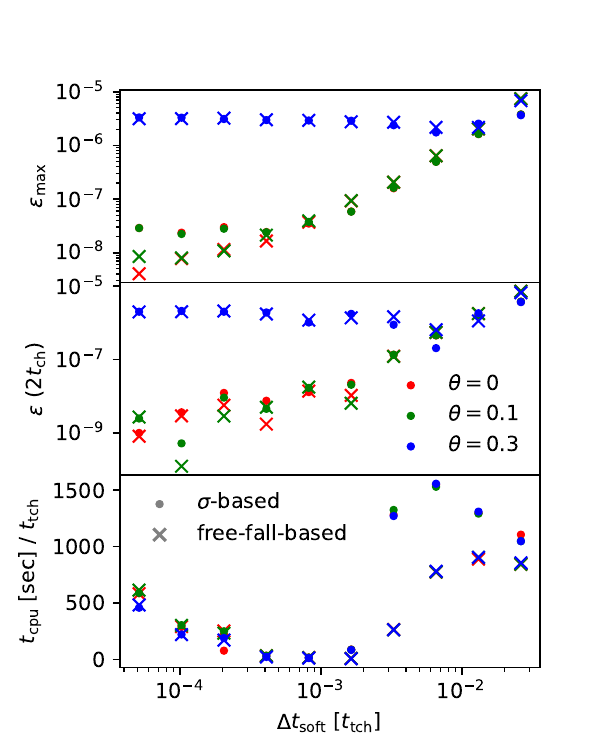}
    \caption{The upper and middle panels show the maximum relative cumulative energy error $\eerrmax$ and the final error at $2~\tcrh$, while the lower panel displays the wallclock time $\tcpu$ per $\tcrh$ up to $2~\tcrh$ for spherical cluster simulations with $\rh=1$~pc, $N=10000$ and equal-mass objects, using both $\sigma$-based and free-fall-based criteria. Results are compared for three opening angles $\theta$.}
    \label{fig:eerr-tcpu-dts-theta-varyN}
\end{figure}

We examine how the opening angle $\theta$ affects the force calculation in the PT method. As $\theta$ increases, computational cost decreases, but accuracy declines. When $\theta$ increases from 0.2 to 0.8, the minimum $\eerrmax$ rises from $10^{-8}$ to $10^{-5}$ (see Figure~2 in \cite{Iwasawa2015}). To balance accurate calculation of angular and linear momentum, as well as relaxation processes, with computational efficiency, $\theta$ should satisfy $0.3\le\theta\le1$ \citep{Hernquist1987}, and thus the default value in \textsc{petar} is set to 0.3. 

We tested $\theta =$ 0.1 and 0.3 alongside $\theta=0$ in the $N=10000$, equal-mass models from Section~\ref{sec:errdt}.
Results are shown in Figure~\ref{fig:eerr-tcpu-dts-theta-varyN}. 
For both $\eerrmax$ and $\eerr(2\tcrh)$, when $\dts>1.3\times10^{-2}~\tcrh$, the values are nearly identical. 
For small $\dts$, $\theta=0.1$ yields $\eerrmax$ nearly identical to $\theta=0$, whereas $\theta=0.3$ produces much lager, $\dts$-insensitive errors above $10^{-6}$. 
Thus, to achieve errors below $10^{-6}$, $\theta = 0.1$ is a safe choice.

Comparing the two criteria, errors differ little for $\theta=0.3$, indicating that the tree-approximation error dominates. The case $\theta=0.1$ shows a similar error behavior to $\theta=0$. 
To highlight the difference between the two criteria, we adopt $\theta = 0.1 $ in subsequent comparisons for various stellar systems.

There is a hidden factor that may affect the $\theta$ comparison: \textsc{fdps} employs a modified PT method \citep{Barnes1990} to improve parallel performance: it forms groups of nearby particles (including adjacent cells) and builds a single interaction list per group rather than per particle. 
Experiments show that the group size affects accuracy. We use 512 to optimize performance on the platform with the AVX2 SIMD instruction set. Increasing the size can reduce errors for \( \theta = 0.3 \). With a group size of 1024 and \( N = 1000 \), all particles fall into one group, making the PT method equivalent to the PP method with second-order leapfrog—hence no difference for any \( \theta \).
For small group sizes, errors for $\theta=0.1$ may decouple from those for $\theta=0$. To ensure $\theta=0.1$ actually uses the tree approximation while maintaining reasonable performance, we use a group size of 512 in all $N=1000$ simulations and 1024 for all simulations with larger $N$.

We also investigated the influence of other parameters listed in Table~\ref{tab:iwasawacompare}.
Because $A_0$ and $\dthm$ affect Hermite timestep estimates, we performed additional simulations using the values from \cite{Iwasawa2015} and observed no change in errors.

In summary, the free-fall-based criterion provides higher accuracy for low-$\sigma$ systems in virial equilibrium, such as $N=1000$ with $\rh=1$~pc, while the $\sigma$-based criterion performs better for high-$\sigma$ systems, such as $N=100000$ with $\rh=1$~pc. Starting from Section~\ref{sec:imf}, we therefore focus on comparing the two criteria for $N=1000$ systems to examine the effects of the IMF, primordial binaries, and non-sphericial or non-virial initial conditions.

\subsubsection{Performance}

The lower panel of Figure~\ref{fig:eerr-tcpu-dts-theta-varyN} compares the wall clock time $\tcpu$ per $\tcrh$ using a single CPU core for each $N=10000$ simulation. For each simulation, 16 OpenMP threads are used and MPI parallelization is switched off. As $\dts$ increases, $\tcpu$ initially decreases, reaching a minimum at $\dts \approx 8.4\times10^{-3}~\tcrh$, and then rises. The initial decrease occurs because larger $\dts$ reduces the number of PT force calculations per $\tcrh$. However, further increases in $\dts$ require larger changeover radii for both criteria. When these radii surpass a threshold, most objects merge into a single large subcluster, where interactions are calculated using the $O(N^2)$ PP method, becoming the main computational bottleneck. In this regime ($\dts > 4.1\times10^{-3}~\tcrh$), $\eerrmax$ also increases with $\dts$, indicating that the \pppt~method becomes unsuitable.

For a given $\dts$, the two criteria show similar performance when $\dts \le 1.6\times10^{-3}~\tcrh$. As seen in Figure~\ref{fig:rin-dts-equalmass-rh1}, the free-fall-based criterion yields larger changeover radii than the $\sigma$-based criterion below $\dts=8.3\times10^{-4}~\tcrh$, and smaller radii above this value. Correspondingly, $\tcpu$ is slightly higher below the transition $\dts$ and lower above it.
When $\dts > 1.6 \times10^{-3}~\tcrh$, the calculation is dominated by $O(N^2)$ PP interactions within a single large subcluster, and the $\sigma$-criterion is much slower because its larger $\rin$ results in more subcluster members.

Larger $\theta$ values result in faster computation only for small $\dts$, though the effect is minor due to the small $N$. The timing difference between $\theta=0$ and 0.3 is slightly larger than between the two criteria. Overall, performance depends more on the choice of $\dts$ than on the criterion or $\theta$.
Because of this, although the criterion lets us derive $\rin$ and $\dts$ from each other, in the following comparisons for various stellar systems, we fix $\dts$ and determine $\rin$ to ensure comparable performance between the two criteria. 

\begin{figure}
    \centering
    \includegraphics[width=\linewidth]{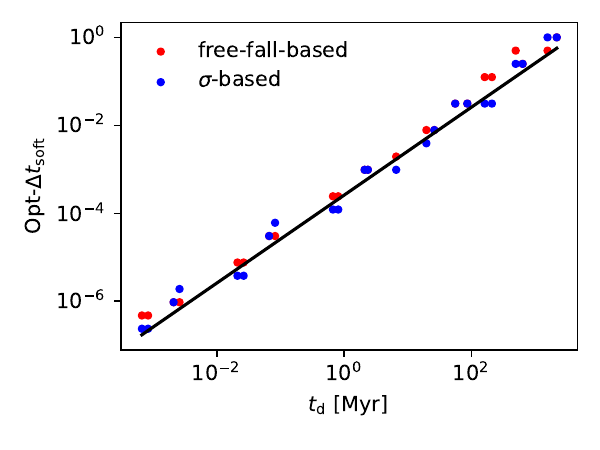}
    \caption{Optimized $\dts$ for peak computing performance versus $\td$ in Plummer stellar systems accross various $N$, $\rh$, object masses, and two criteria. The black line shows a linear fit with the coefficient $5\times10^{-5}$.}
    \label{fig:optdts}
\end{figure}

The optimal $\dts$ (Opt-$\dts$) for peak computing performance depends on the stellar system. 
We evaluate Opt-$\dts$ for $N=1000,10000,100000$ and $\rh=0.01, 0.1, 1, 10, 100$~pc in Plummer clusters with equal-mass stars and the Kroupa IMF, using $\theta=0.1$ and both criteria. The results in Figure~\ref{fig:optdts} show that, despite parameter-dependent differences, Opt-$\dts$ scales with
\begin{equation}
    \td = \frac{G M}{\sigma^3},
\end{equation}
which is equivalent to $\tcr$ in virial equilibrium. A slightly adjusted linear fit gives Opt-$\dts \approx 2.6\times10^{-4}\td$. 

\subsection{Spherical star clusters with IMF}
\label{sec:imf}

\begin{figure}
    \centering
    \includegraphics[width=\linewidth]{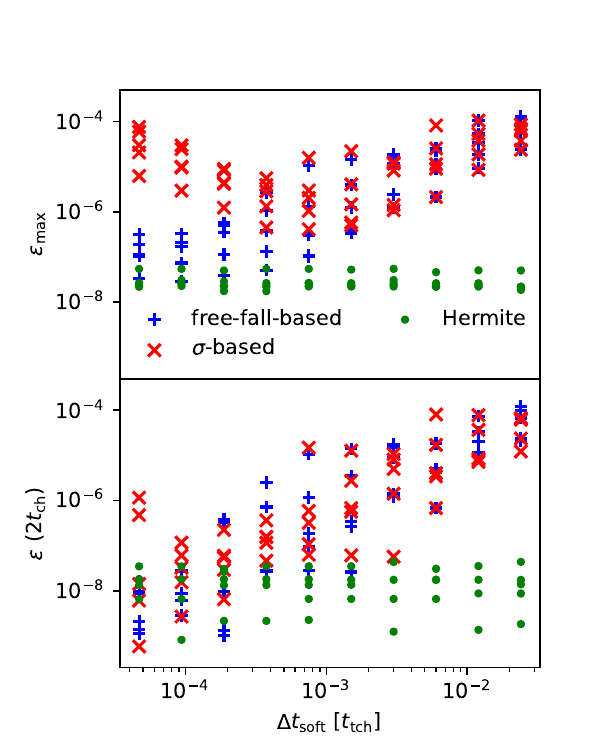}
    \caption{As in Figure~\ref{fig:eerr-t-equalmass-rh1}, but for star clusters with an IMF.}
    \label{fig:eerrmax-dts-optimf-rh1}
\end{figure}

We investigate how the two criteria behave when introduce multi-mass objects in star cluster simulations. 
We conduct star cluster simulations similar to those in Section~\ref{sec:equalmass} with $N=1000$ except that the mass of objects are sampled from the \cite{Kroupa2001} IMF.
To minimize the significant variability in dynamical evolution caused by large fluctuations in the number of massive stars when randomly sampling the IMF across different seeds \citep{Wang2021}, we use optimal sampling \citep{Yan2017}, which fixes both the number of massive stars and the maximum stellar mass.
We adopt the multi-mass free-fall-based criterion (Equation~\ref{eq:imfcrit}). 

Results comparing the $\sigma$-based, free-fall-based, and pure Hermite methods are presented in Figure~\ref{fig:eerrmax-dts-optimf-rh1}. Compared to the case of equal mass in Figure~\ref{fig:eerrmax-dts-equalmass-rh1}, IMF simulations show a greater scatter between random seeds, but the overall trend remains similar.

\subsection{Spherical star clusters with primordial binaries}
\label{sec:binary}

In star clusters with many binaries, the $\sigma$-based criterion faces a challenge of accurately measuring $\sigma$. 
Tight binaries have much higher relative velocities than most single stars. 
If all binaries are resolved when calculating $\sigma$, the value becomes much larger than in clusters without binaries, leading the $\sigma$-based criterion to set an excessively large $\rin$ and reducing simulation efficiency. 
Conversely, if all binaries are treated as unresolved and only their center-of-mass velocities are used, $\sigma$ may be underestimated, resulting in a $\rin$ that is too small to accurately model some wide binaries for the PT part.

\begin{table*}[]
    \centering
    \caption{The values of $\rin$, $\sigma$ (at 0 Myr) and $\tcpu$ (per Myr) for star clusters with primordial binaries using two criteria.}
    \label{tab:rinbinary}
    \begin{tabular}{cc|cc|ccc|ccc}
\hline
$r_{\mathrm{h}}$ & $\Delta t_{\mathrm{soft}}$ & $r_{\mathrm{in}}$ & $t_{\mathrm{cpu}}$& $\sigma$ & $r_{\mathrm{in}}$ & $t_{\mathrm{cpu}}$ & $\sigma$ & $r_{\mathrm{in}}$ & $t_{\mathrm{cpu}}$ \\
\text{[pc]} & \text{[Myr]} & \text{[pc]} & \text{[sec]} & \text{[pc/Myr]} & \text{[pc]} & \text{[sec]} & \text{[pc/Myr]} & \text{[pc]} & \text{[sec]} \\
 & & \multicolumn{2}{c|}{free-fall-based}  & \multicolumn{3}{c|}{unresolved binaries, $\sigma$-based} & \multicolumn{3}{c}{resolved binaries, $\sigma$-based} \\
\hline
1 & 0.00049 & 0.005 & $3.2 \times 10^{2}$ & 0.92 & 0.0022 & $1.4 \times 10^{2}$ & 31 & 0.076 & $8.7 \times 10^{3}$\\
100 & 0.0039 & 0.02 & $2.1 \times 10^{0}$ & 0.11 & 0.0021 & $4.6 \times 10^{0}$ & 31 & 0.61 & $2.7 \times 10^{0}$\\
\hline
    \end{tabular}
\end{table*}

To illustrate this issue, we run simulations for a spherical star cluster with $\rh = 1$, $N=1000$ and $100$~pc, using the Kroupa IMF and the Kroupa primordial binary model \citep{Kroupa1995a,Kroupa1995b,Sana2012,Belloni2017}, which assumes all stars initially form in binaries with a broad range of periods and eccentricities. This creates an extreme scenario to test the criteria’s performance. To keep sufficient accuracy for binary integrations, we apply the \textsc{sdar} method with $\rg = 0.8~\rin$. 
Applying the two best-performing $\dts$ criteria, the initial $\sigma$, $\rin$ and $\tcpu$ values and are listed in Table~\ref{tab:rinbinary}.

When binaries are resolved, the initial $\sigma \approx 31$~pc/Myr for both $\rh$, driven by tight binaries rather than the whole cluster. The $\sigma$-based $\rin$ is 15–30 times larger than the free-fall-based value, resulting in much slower simulations for dense clusters. For $\rh=1$~pc, the $\sigma$-based approach is about 35 times slower, as roughly 80\% of objects merge into a single large subcluster requiring the expensive PP method. For $\rh=100$~pc, the performances are similar, as the low cluster density prevents the formation of a large subcluster.

\begin{figure}
    \centering
    \includegraphics[width=\linewidth]{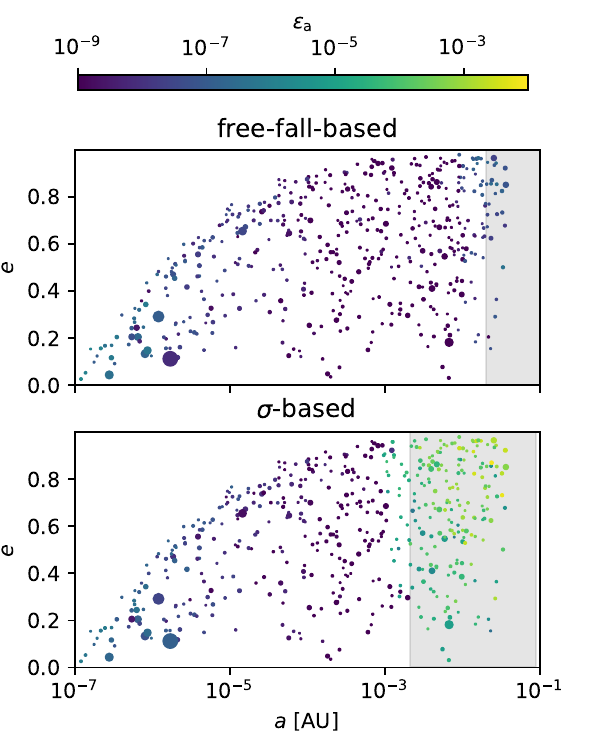}
    \caption{Relative error in $a$ ($\erra$, shown as colors) after approximately $1.2$ Myr for spherical star clusters with Kroupa-model primordial binaries, comparing free-fall-based and $\sigma$-based criteria (including unresolved binaries). Each point shows the $a$ and $e$ of individual binaries, with point size indicating binary mass. The grey region marks the changeover region bounded by the minimum $\rinij$ and maximum $\routij$.}
    \label{fig:semi-ecc-rh100}
\end{figure}

With unresolved binaries, $\sigma$ reflects the overall cluster dynamics. For $\rh=1$~pc, the $\sigma$-based $\rin$ is about half the free-fall-based value and remains acceptable, but for $\rh=100$~pc, it is ten times smaller, risking low accuracy for wide binaries.

To demonstrate this, we analyze the accuracy of integrating binaries in the cluster with $\rh=100$~pc for about $1.2$~Myr—twice the maximum binary period. In such low-density conditions, binaries experience minimal external perturbations and their semi-major axes $a$ should remain constant. Thus, the relative change $\erra = \left| \left[a(t)-a(0)\right]/a(0) \right|$ effectively meansures integration error. Figure~\ref{fig:semi-ecc-rh100} compares $\erra$ for individual binaries under the free-fall-based and $\sigma$-based criteria.

For binaries with $a>10^{-3}$~AU in the changeover region of the $\sigma$-based criterion, errors are up to two orders of magnitude higher than in the free-fall-based case. When $a<\rin$, the \textsc{sdar} method yields accurate results, but in the changeover region, mixing Hermite and leapfrog integrators causes relatively large errors, up to $10^{-2}$. 
In dense clusters, such wide binaries are quickly disrupted by encounters, making low-accuracy methods acceptable. In low-density clusters, however, wide binaries can persist, so these errors become significant.

For similar $a$, errors grow with increasing $e$ because high-eccentricity orbits are harder to handle, as shown in Figure~\ref{fig:eerrmax-dts-ecc}.

These results show that, because the free-fall-based criterion is independent of cluster density, it offers a better balance of performance and accuracy than the $\sigma$-based criterion in clusters with many binaries, regardless of density.

\subsection{Spherical star clusters with different virial ratios}

During the early star-forming phase, star clusters are often not in virial equilibrium. The virial ratio $Q = E_{\mathrm{k}}/\left|E_{\mathrm{V}}\right|$, with $E_{\mathrm{k}}$ and $E_{\mathrm{V}}$ as the total kinetic and potential energies, indicates the system’s virial state. 
Clusters collapse when $Q<0.5$ (subvirial) and expand when $Q>0.5$ (supervirial). 
In collapsing systems, the initial velocity dispersion $\sigma$ is low and quickly rises as the cluster re-equilibrates. If the $\sigma$-based criterion uses the initial $\sigma$ to set $\rin$, it may underestimate $\rin$ in these cases, whereas the free-fall-based criterion, being independent of $\sigma$, is less affected.

\begin{figure}
    \centering
    \includegraphics[width=\linewidth]{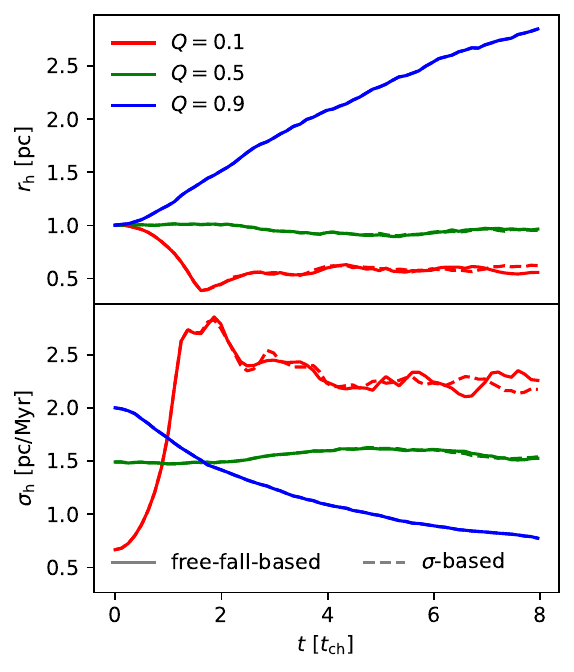}
    \caption{Evolution of half-mass radius $\rh$ (top) and half-mass velocity dispersion $\sh$ (bottom) for spherical star clusters with equal-mass objects and different virial ratios $Q$. The two criteria are indicated by different line styles.}
    \label{fig:rh-sigma-t-q}
\end{figure}

To investigate how $Q$ influences integration accuracy, we simulate spherical star clusters ($\rh=1$~pc and $N=1000$) up to about $8~\tcrh$ with equal-mass objects (3.8~Myr) and the Kroupa IMF (5.2~Myr) for $Q=0.1$, 0.5, and 0.9, corresponding to subvirial, virial equilibrium, and supervirial cases. 
Five simulations with different random seeds are run for each $Q$, applying both criteria for comparison. 
We use the optimized $\dts=0.0009765625$, which yields best computing performance (Opt-$\dts$) for all $Q$.
In the equal-mass case, the $\sigma$-based criterion yields $\rin \approx 0.0020$, 0.0045, and 0.0060~pc for $Q=0.1$, 0.5, and 0.9, respectively, all smaller than the free-fall-based $\rin$ of $\approx 0.0078$~pc.

For the subvirial simulations, to investigate whether reducing $\dts$ and increasing $\ns$ improves accuracy, we add runs with $\dts$ reduced by a factor of 8 (Div8-$\dts$) for both the equal-mass and IMF cases, and by a factor of 32 (Div32-$\dts$) for the IMF case. For the free-fall-based criterion in the IMF case, we also include Opt- and Div8- simulations with $\ns$ increased to 256 (Ns256) for comparison.

Figure~\ref{fig:rh-sigma-t-q} shows the evolution of $\rh$ and half-mass velocity dispersion $\sh$ using Opt-$\dts$, comparing equal-mass simulations with one random seed are. The IMF case is similar and therefore omitted.
As expected, in subvirial clusters $\rh$ decreases and $\sh$ increases rapidly; at virial equilibrium both remain stable; in supervirial conditions $\rh$ expands while $\sh$ declines. 
Both criteria yield similar results, except for subvirial clusters, where differences become  noticeable after $2~\tcrh$ due to larger errors discussed below.

\begin{figure}
    \centering
    \includegraphics[width=\linewidth]{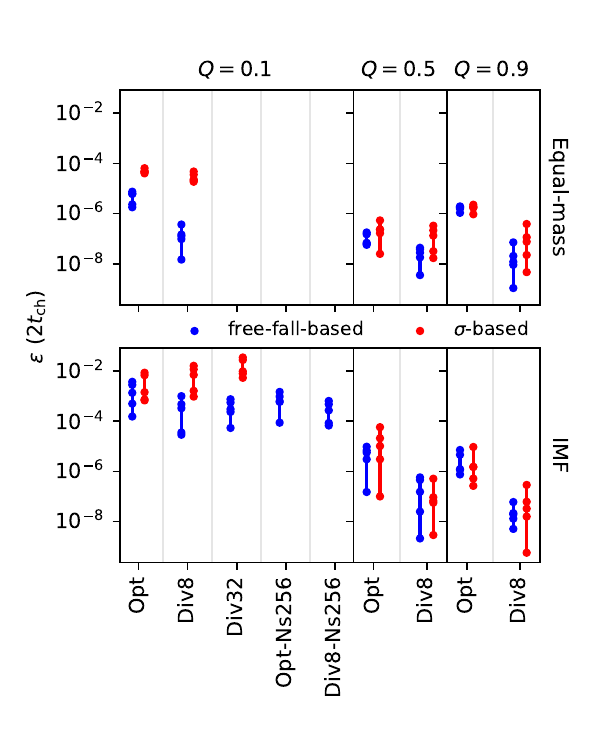}
    \caption{Final relative cumulative energy error $\eerr$ at $8~\tcrh$ versus $\dts$ for different virial ratios of spherical clusters with equal-mass and a Kroupa IMF. The free-fall-based criterion and $\sigma$-based criterion are compared. For $Q=0.1$, IMF simulations with additional $\dts$ values and varying $\ns$ are also compared. For each parameter pair, errors from simulations with different random seeds are shown as markers, with mininum and maximum values values connected by lines for clarity.}
    \label{fig:eerrf-q}
\end{figure}

Figure~\ref{fig:eerrf-q} shows the final $\eerr$ at $8~\tcrh$ for all seeds, comparing the equal-mass and Kroupa IMF cases. The subvirial ($Q=0.1$) case exhibits the highest $\eerr$, consistent with its large post-collapse $\sh$. 
In the equal-mass case, the supervirial ($Q=0.9$) case has higher $\eerr$ than the equilibrium ($Q=0.5$) case, likely due to a larger initial $\sh$. 
However, for the IMF case, this trend reverses, indicating that errors from the multi-mass effect outweigh those from the initial $\sh$. Notably, for the same $Q$, supervirial and equilibrium cases show similar errors in both equal-mass and IMF runs, while for the subvirial case, the IMF yields much larger errors than the equal-mass case.

With Opt-$\dts$, the free-fall-based criterion generally yields smaller errors than the $\sigma$-based criterion. In subvirial equal-mass runs, the $\sigma$-based criterion gives much larger errors due to its underestimated $\rin$ (0.0020 for $\sigma$-based versus 0.0078~pc for free-fall-based). 
After collapse, $\sh$ increases to four times its initial value as shown in Figure~\ref{fig:rh-sigma-t-q}. If $\rin$ is recalculated using this peak $\sh$, it increases by a factor of four, aligning with the free-fall-based value.
This suggests that redetermining $\rin$ after collapse could improve the $\sigma$-based criterion's energy error.

For the equal-mass case, Div8-$\dts$ significantly reduces errors for the free-fall-based criterion, making the subvirial error comparable to the equilibrium case. However, the improvement is not evident for the $\sigma$-based criterion, except in the supervirial case.
For the IMF case, the free-fall-based criterion again shows better accuracy with Div8-$\dts$, but the improvement in the subvirial condition is limited, with error above $10^{-5}$.
Interestingly, the $\sigma$-based criterion with Div8-$\dts$ shows better improvement in the equilibrium and supervirial cases but performs worse in the subvirial case.

Increasing $\ns$ for the free-fall-based criterion (Opt-Ns256-$\dts$) allows for a larger $\rin$ at Opt-$\dts$ and gives errors comparable to Div8-$\dts$, but Div32-$\dts$ and Div8-Ns256-$\dts$ offer no further improvement. This indicates the difficulty of achieving high energy conservation under subvirial conditions with an IMF.

\begin{figure}
    \centering
    \includegraphics[width=\linewidth]{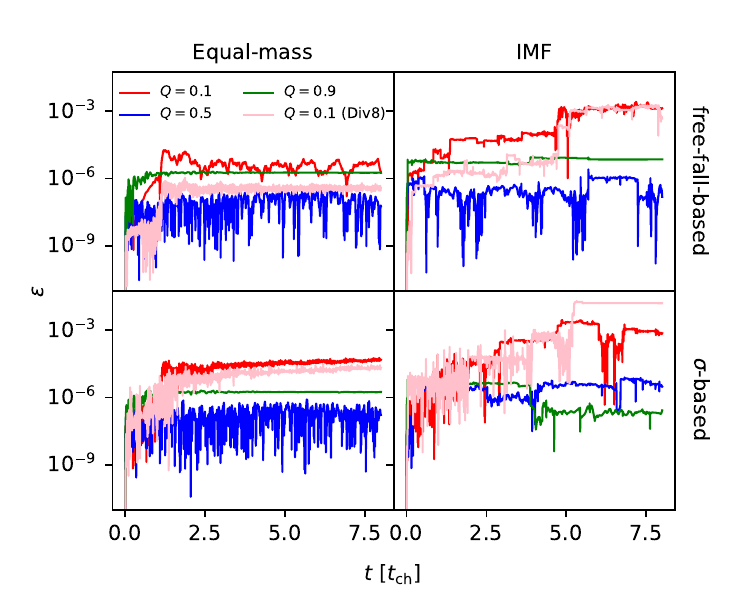}
    \caption{Relative cumulative energy error $\eerr$ over time for spherical star clusters with equal-mass objects (left) and Kroupa IMF (right), for different virial ratios $Q$, comparing two criteria (top and bottom) for a single random seed. Colors represent different $Q$, with $Q=0.1$ including a Div8-$\dts$ (pink) for comparison.}
    \label{fig:eerr-t-q}
\end{figure}

To investigate how $\eerr$ grows with different $Q$ and criteria, Figure~\ref{fig:eerr-t-q} shows $\eerr$ evolution for a single seed with equal-mass objects and the Kroupa IMF. In the subvirial case, $\eerr$ with Opt-$\dts$ rises rapidly during the first 2~$\tcrh$ of collapse. With Div8-$\dts$, the free-fall-based $\eerr$ for equal-mass objects matches that of the equilibrium case, while the $\sigma$-based criterion does not. For the IMF case, the free-fall-based criterion yields a lower energy error before $5~\tcrh$, but exhibits a sharp increase afterward. In contrast, the $\sigma$-based criterion has a similar error level to Opt-$\dts$ before $5~\tcrh$. Thus, although the free-fall-based criterion shows a large error in the end, it provides a lower average energy error overall. Therefore, only the free-fall-based criterion can reduce the energy error as $\dts$ decreases.

\subsection{Fractal star clusters}

\begin{figure*}
    \centering
    \includegraphics[width=\linewidth]{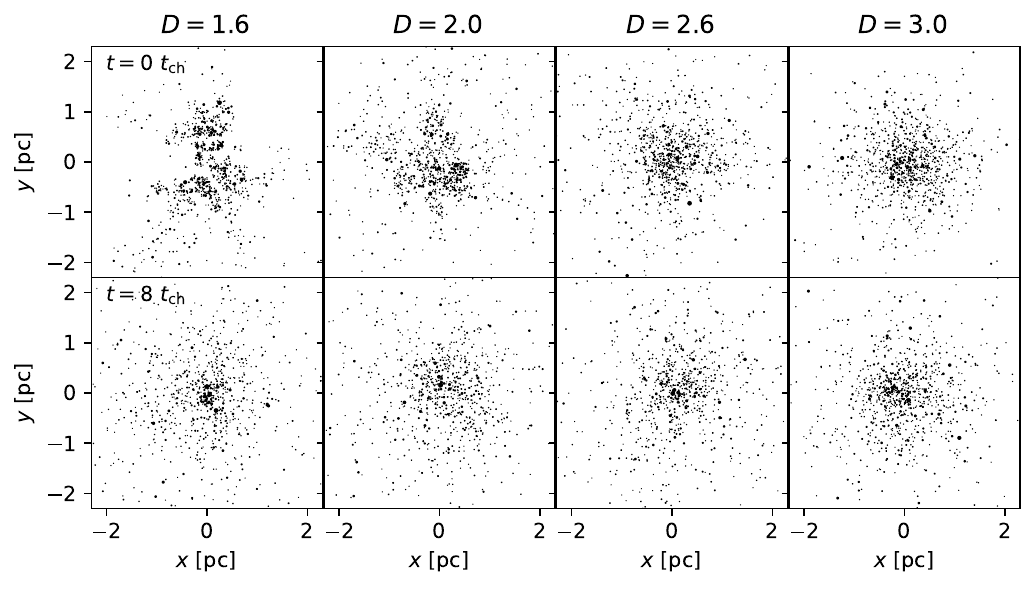}
    \caption{Initial (top) and final (bottom) stellar distributions for fractal clusters, with point size indicating object mass (Kroupa IMF).}
    \label{fig:frac-xy-imf}
\end{figure*}

During the gas-embedded phase, newly formed star clusters begin with a clumpy structure and substructures rather than a spherical shape. To model this, we use \textsc{mcluster} to generate fractal star clusters with $N=1000$, equal-mass objects and the Kroupa IMF, following \cite{Goodwin2004}, and adopt fractal dimension parameters $D=1.6$, $2.0$, $2.6$, and $3.0$ to span the range from highly clumpy to sphercial structures. For each $D$, five models with different random seeds are created. The initial effective $\tcrh$ is approximately 0.47 Myr for the equal-mass case and 0.65 Myr for the IMF case.

Figure~\ref{fig:frac-xy-imf} shows the initial and final morphology for one seed with the IMF after $8~\tcrh$ (about $\sim5.2$~Myr) of evolution. The $D=1.6$ clusters are the most clumpy initially, but after $8~\tcrh$, all cases become nearly spherically symmetric.

\begin{figure}
    \centering
    \includegraphics[width=\linewidth]{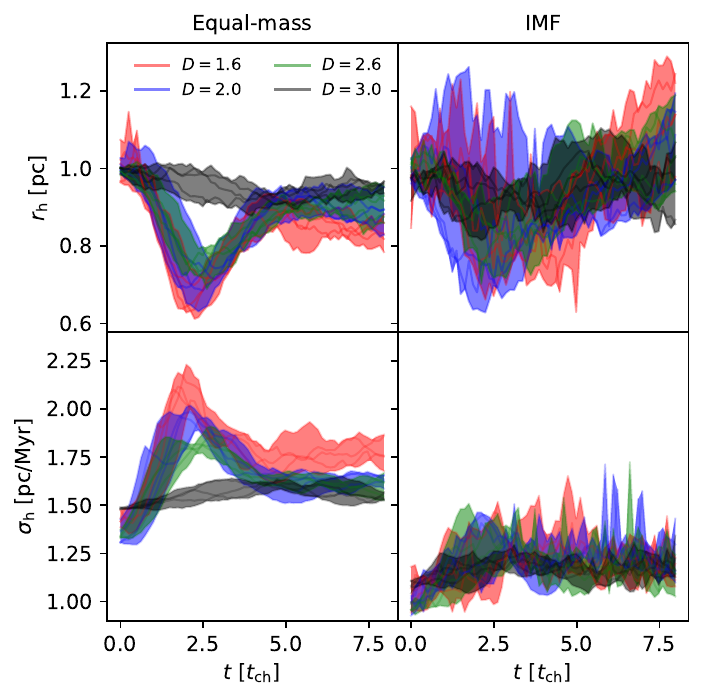}
    \caption{Evolution of half-mass radius $\rh$ (top) and velocity dispersion $\sh$ (bottom) for fractal star clusters with equal-mass objects (left) and the Kroupa IMF (right). For each color (denoting $D$), each curve shows one of five simulations with different random seeds, and the shaded region spans the mininum-maximum range for clarity.}
    \label{fig:rh-sigma-t-frac}
\end{figure}

Figure~\ref{fig:rh-sigma-t-frac} shows the evolution of $\rh$ and $\sh$ for both equal-mass and IMF cases. For equal-mass clusters, all fractal models except the spherical $D=3.0$ cluster initially collapse, then expand to a similar $\rh$ as the $D=3.0$ case, with $\sh$ first increasing and then decreasing. Different random seeds yield similar results. 
This pattern reflects the collapse of clumpy structures followed by expansion toward global virial equilibrium.

In the IMF case, the evolution is more stochastic for a given $D$ and lacks the clear trends seen in the equal-mass case. For $D=1.6$ and 2.0, $\rh$ increases in some runs and decreases in others, while $\sh$ varies less across different seeds and $D$ values. The wide mass range and small number of massive stars make their positions highly stochastic, driving the overall variability in the evolution.

Because initial $\sh$ values are similar for different $D$, the estimated $\rin$ from the $\sigma$-based criterion is also similar. At the optimal time step $\dts=0.0009765625$~Myr for computing performance, the $\sigma$-based criterion yields $\rin$ of $0.0063$–$0.0069$~pc for the equal-mass case and $0.0045$–$0.0052$~pc for the IMF case. The free-fall-based criterion gives $\rin$ of $0.0078$~pc (equal-mass) and $0.0096$~pc (IMF), which are not significantly different.

\begin{figure}
    \centering
    \includegraphics[width=\linewidth]{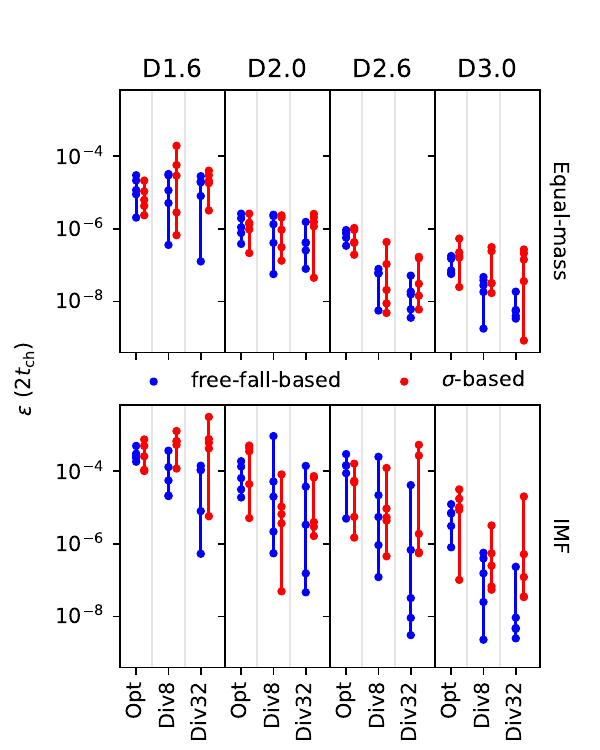}
    \caption{Final relative cumulative energy error at $8~\tcrh$ versus fractal parameter $D$ for equal-mass and Kroupa IMF cases, comparing both criteria. Plotting style follows Figure~\ref{fig:eerrf-q}.}
    \label{fig:eerrf-frac}
\end{figure}

Figure~\ref{fig:eerrf-frac} shows $\eerr$ at $8~\tcrh$ for different $D$ and $\dts$ values. Generally, $\eerr$ decreases as $D$ increases, suggesting that fractal initial conditions lead to more stochastic strong interactions. Both equal-mass and IMF cases show similar trend and the IMF case have systematically larger $\eerr$ for all $D$.

Comparing the two criteria, the free-fall-based approach generally outperforms the $\sigma$-based one. For smaller $\dts$, errors show no clear trend with the \(\sigma\)-based criterion, while they typically decrease with the free-fall-based criterion, albeit with large scatter.

\section{Discussion}
\label{sec:discussion}

We introduce the free-fall-based criterion for the \pppt~scheme, but this approach can be extended to other hybrid methods that combine accurate short-range and fast approximate long-range integrators. The $\ns$ value may vary depending on the choice of integrator. For instance, using higher-order symplectic methods for the PT part or alternative Hamiltonian splittings may reduce the required $\ns$ when $\theta$ is small.

For the \pppt~method, neither the free-fall-based nor the $\sigma$-based criterion matches the accuracy of Hermite with adaptive time steps at optimized $\dts$, as shown in Figures~\ref{fig:eerrmax-dts-equalmass-rh1} and~\ref{fig:eerr-tcpu-dts-theta-varyN}. This indicates that increasing the step control parameter $\eta$ can enhance computing performance for Hermite (PP part) at the cost of some accuracy when strict energy conservation is not essential.

In addition, both PT and PP integrations use double-precision (64-bit) in our simulations, while the default in \textsc{petar} assigns single precision (32-bit) to PT and double precision to PP for greater efficiency. Using single precision for PT limits $\eerr$ to about $10^{-6}$ due to roundoff errors, making Hermite's higher accuracy of $10^{-8}$ to $10^{-7}$ unattainable for the PT part. Thus, increasing $\eta$ to align Hermite’s accuracy with the $10^{-6}$ threshold has minimal impact on global $\eerr$. As shown by \cite{Boekholt2015}, this accuracy is sufficient for statistical equivalence with more precise simulations for triple systems, and most studies of collisional star clusters adopt an energy error limit of $10^{-6}$.

Alternatively, when high accuracy is required for simulations, the PT component's precision can be enhanced by using a more accurate integrator in place of leapfrog rather than reducing $\dts$. This approach can boost performance and may resolve accuracy issues in modeling fractal and subvirial stellar systems.

We have examined both criteria for various stellar systems. For star clusters with a Plummer model and equal-mass distribution, we investigate the different $N$ up to $10^5$. In fact, the \pppt~method offers greater benefits for larger $N$, improving both performance—thanks to $O(N \log N)$ scaling—and accuracy. As $N$ increases, the two-body relaxation timescale increases and the PT method’s force accuracy improves because the long-range gravitational potential becomes smoother, resulting in more accurate long-range forces \citep{Hernquist1987}. Therefore, we have explored more challenging scenarios for the \pppt~method.

The two criteria are compared with fixed parameters in most simulations: $\ns=64$ for the free-fall-based criterion and $\alpha=0.2$ for the $\sigma$-based criterion. While increasing these parameters can improve accuracy in challenging cases, as seen in the subvirial simulations (Figure~\ref{fig:eerrf-q}), we did not explore this further, focusing instead on how each criterion performs across different stellar systems with a fixed parameter set.

\section{Conclusion}
\label{sec:conclusion}

We propose a free-fall-based criterion (Equation~\ref{eq:crit} and \ref{eq:imfcrit}) to set the changeover radii and the global tree time step $\dts$ for the \pppt~$N$-body algorithm, ensuring sufficient accuracy when switching between PT and PP methods.
Unlike the $\sigma$-based criterion (Equation~\ref{eq:sigmabase}), this criterion is easier to evaluate, requires no measurement of $\sigma$, and thus is more versatile, particularly for systems with binaries or for low-$\sigma$, subvirial, and irregular systems where $\sigma$ is difficult to accurately determine. 

We test both criteria across a range of astrophysically relevant stellar systems, including spherical star clusters with equal-mass stars, the Kroupa IMF, primordial binaries, various virial ratios, and fractal star clusters. Overall, we find that the free-fall-based criterion with $\ns=64$ yields better accuracy and error scaling with respect to $\dts$ than the $\sigma$-based criterion with $\alpha=0.2$ for $N=1000$.
For high-$\sigma$ systems, such as spherical star clusters with $N=10000$ and $\rh=1$~pc, or more massive systems like GCs and NSCs, the $\sigma$-based criterion is more accurate due to larger $\rin$ for a given $\dts$. In a virial equilibrium system, the transition timestep at which the two criteria yield the same \(\rin\) satisfies the simple relation \(\dts/\tcrh \propto 1/N\). In practice, the two criteria can be combined to cover the full range of densities. 

For spherical star clusters with equal-mass stars, both criteria yield similar $\eerr$ when $\dts$ exceeds the optimal value (Figure~\ref{fig:eerrmax-dts-equalmass-rh1} and~\ref{fig:eerr-tcpu-dts-theta-varyN}). However, at smaller $\dts$, especially for low-$\sigma$ system like $N=1000$, the free-fall-based criterion achieves higher accuracy, while the $\sigma$-based criterion worsens, as its changeover radii are too small to capture close encounters (Figure~\ref{fig:eerr-t-equalmass-rh1}). For $N\ge10^5$ with $\rh=1$~pc, the $\sigma$-based criterion has better accuracy for most acceptable $\dts$. 

For opening angles $\theta \le 0.1$, accuracy is nearly independent of $\theta$ (Figure~\ref{fig:eerr-tcpu-dts-theta-varyN}), making the PT method indistinguishable from the direct $N$-body approach ($\theta=0$) in this range, consistent with \cite{Hernquist1987}. However, this may change when $N$ or the PT group size differs. 

For spherical star clusters with the Kroupa IMF, both criteria yield higher $\eerr$ at the same $\dts$ (Figure~\ref{fig:eerrmax-dts-optimf-rh1}). The wider stellar mass range makes the accuracy harder to achieve and increases the stochastic variations in $\eerr$, though the overall trend remains similar to the case of equal mass.

For spherical clusters with the Kroupa IMF and  primordial binary model, the free-fall-based criterion shows a significant advantage in ensuring the integration accuracy of wide binaries in a low-density clusters (Figure~\ref{fig:semi-ecc-rh100}). In this case, $\sigma$ can have a large variation depending on whether binaries are resolved. In the unresolved case, the $\rin$ is too small so that wide binaries with $a>10^{-3}$~AU are not included in the accurate \textsc{sdar} method, thus having large errors after only 1.2 Myr of simulations. 

For spherical clusters with subvirial initial conditions, the collapse poses challenges for energy conservation with both criteria. Reducing $\dts$ improves errors for the free-fall-based criterion, but not for the $\sigma$-based criterion, as the initially low $\sigma$ leads to an underestimated $\rin$.

Fractal initial conditions are the most challenging for both criteria. Reducing \(\dts\) can improve the free-fall-based criterion, but gains are stochastic, while the \(\sigma\)-based criterion benefits little. This reflects the complex evolution of clumpy structures toward virial equilibrium.

Future work should examine whether replacing leapfrog in the PT part with a more accurate integrator improves accuracy under subvirial and fractal conditions.

\begin{acknowledgments}
LW thanks the support from the National Natural Science Foundation of China through grant 12233013, 12573041 and 21BAA00619, the Fundamental Research Funds for the Central Universities, Sun Yat-sen University (2025QNPY04), the High-level Youth Talent Project (Provincial Financial Allocation) through the grant 2023HYSPT0706.  DMH acknowledges support from the National Science and Technology Council (NSTC) in Taiwan through grant 114-2112-M-003 -020 -MY3.
\end{acknowledgments}

\begin{contribution}

LW was responsible for the research concept, performing the simulations, data analysis, and writing and submitting the manuscript.
DMH co-developed the research concept, contributed to formula derivation and discussions, and edited the manuscript.
ZZ and WH contributed to the discussions.



\end{contribution}

%

\software{numpy (\citealp{harris2020array}),
          matplotlib (\citealp{Hunter:2007}),
          \textsc{sdar} (\citealp{Wang2020a}, https://github.com/lwang-astro/SDAR),
          \textsc{petar} (\citealp{Wang2020b}, https://github.com/lwang-astro/PeTar),
          \textsc{fpds} (\citealp{Iwasawa2016},
          https://github.com/FDPS/FDPS),
          \textsc{mcluster} (\citealp{Kuepper2011}, https://github.com/lwang-astro/mcluster)
          }


\appendix

\bibliography{paper}{}
\bibliographystyle{aasjournalv7}



\end{document}